\documentclass[]{aa501}
\usepackage{epsfig}
\usepackage{times}
\newcommand{\sub}[1]{_{\rm #1}}
\newcommand{\HII}{H{\sc II} }
\newcommand{\changed}{ }

\begin{document}


\title{Massive core parameters from 
spatially unresolved multi-line observations}
\titlerunning{Massive core parameters}

\author{V.~Ossenkopf, C.~Trojan\and{} J.~Stutzki}
\authorrunning{Ossenkopf et al.}

\institute{
1. Physikalisches Institut, Universit\"at zu K\"oln,
Z\"ulpicher Stra\ss{}e 77, 50937 K\"oln, Federal Republic of Germany}

\date{Received: 15 November 2000; accepted: 18 June 2001}

\abstract{
We present observations of 15 massive cores in three different CS 
transitions from the FCRAO 14m and the KOSMA 3m telescope.
We derive physical parameters of these cores using different 
approaches to the line radiative transfer problem. The local 
radiative transfer approximations fail to provide reliable values 
except for the column densities. A self-consistent explanation of 
the observed line profiles is only possible when taking density 
gradients and an internal turbulent structure of the cores into 
account. The observational data can be fitted 
by a spherically symmetric radiative transfer model including
such gradients and a turbulent clumping.
We find that the observed massive cores are approximately virialised
with a clumpy density profile that decays with a radial exponent of 
about $-1.6$ down to a relatively sharp outer boundary.
\\
We show that a careful analysis of spatially unresolved multi-line
observations using a physical radiative transfer model can provide
values for physical parameters that could be obtained otherwise only 
by direct observations with much higher spatial resolution. This applies
to all quantities directly affecting the line excitation, like the 
mass and size of dense cores. Information on the exact location or 
number of clumps, of course, always 
has to rely on high-resolution observations e.g. from interferometers.
\keywords{ Line: profiles --  Radiative transfer --
        ISM: clouds -- ISM: structure -- Radio lines: ISM }
}
\maketitle

\section{Introduction}

Whereas the average density in molecular clouds falls between about
50 and 1000 cm$^{-3}$, observations of high-dipole-moment
molecules like CS or NH$_3$ reveal cores with densities up to 
$10^6$ cm$^{-3}$. Massive cores are typically somewhat warmer
than their embedding molecular cloud and show sizes of about $1- 3$~pc 
and masses between some ten and some thousand solar masses. 
They appear as relatively bright objects in molecular line maps 
so that they are favoured objects from the viewpoint of the
observations. The observed line widths of $2-15$ km/s are considerably 
larger than thermal, indicating turbulent motions, possibly induced by
outflows from star-forming activity. Many of them are associated
with young OB or T Tauri stars suggesting that massive cores 
are sites of massive and multiple star formation (c.f. \cite{Myers}).
For a better understanding of star formation we have to know 
the physical parameters within these cores, i.e. the geometrical structure,
the density, velocity, and temperature distribution. 

Dense cores are best traced by molecules like CS, HC$_3$N, NH$_3$, and
H$_2$CO characterised by high critical densities even for low transitions.
Throughout this paper we will use the term ``core'' for the dense 
inner part of a cloud which is directly visible in CS, whereas the
rest of the cloud may contribute to the radiative excitation but
is mostly visible in CO rather than in CS. We call smaller substructures 
within the cores ``clumps''. Meta-stable transitions 
of NH$_3$ provide reasonable estimates for the kinetic core temperature 
and the simultaneous observation of several lines from the same 
isotope provide combined information both on the
density and the temperature structure. Here, we  use observations
of the 2--1, 5--4, and 7--6 transitions  of CS and C$^{34}$S obtained
with the FCRAO 14m and the KOSMA 3m telescopes to study
15 massive cores and complement our observations with CS 
data from the literature to derive the physical core parameters.

\mbox{}\cite{Plume} observed about 150 massive cores in CS and C$^{34}$S using
the IRAM 30m telescope. With an escape probability approximation
they derive relatively similar physical parameters for most cores.
We compare their results to the parameters obtained for the
cores from our sample and test to which extent these results
reflect the constraints provided by the set of lines observed,
the limitations of the data analysis or the real physical properties 
of the cores. To study the influence of the data analysis
we compare the escape probability approximation with a self-consistent 
radiative transfer code computing the excitation conditions in an 
inhomogeneous core with internal turbulence.

In Sect. 2 we provide a short overview on the sample and the observations.
Sect. 3 discusses the traditional way to derive the cloud parameters
from the observations. Using the mathematical description of the
radiative transfer problem in Appx.\ A we compute cloud parameters 
in an escape probability approximation. In Sect. 4 the fully 
self-consistent radiative transfer code  from Appx.\ B is used to 
derive the cloud parameters. Sect. 5 compares the resulting data
with parameters obtained from independent observations 
and discusses implications for the physics of massive cores.

\section{Observations}

\subsection{The sample}

The selection of the sample of massive cores was determined by
the need of bright ``standard'' sources for the SWAS satellite which has
a spatial resolution of about 4~arcmin in the frequency range between
490 and 560 GHz. For a comparison to data obtained at similar angular 
resolution we used observations taken with the array receiver of the FCRAO 
14m telescope at 98 GHz (about 1~arcmin resolution) and performed 
complementary observations with the 3m KOSMA telescope at 245 and 343 GHz 
where its beam size is approximately 2~arcmin.

The sources were selected from the SWAS source list (Goldsmith
et al., {\it priv. comm.}) to be observable from the FCRAO and KOSMA
and bright enough to be detectable in a reasonable integration
time. Tab. \ref{tab_sources}  lists the 15 sources selected with 
their central position.

\begin{table}
\caption{Selected source sample with central positions}
\begin{tabular}[h]{lrr}
\multicolumn{1}{c}{Source}
& \multicolumn{1}{c}{ $\alpha_{1950}$ } 
& \multicolumn{1}{c}{ $\delta_{1950}$ } \\ \hline 
W49A & ${19}^{h}:{07}^{m}:{50.0}^{s}$ & ${09}^{\circ}:{01'}:{32.0''}$  \\ 
W33 & ${18}^{h}:{11}^{m}:{19.8}^{s}$ & ${-17}^{\circ}:{56'}:{21.0''}$  \\ 
W51A & ${19}^{h}:{21}^{m}:{25.4}^{s}$  & ${14}^{\circ}:{24'}:{45.0''}$  \\
W3(OH) & ${2}^{h}:{23}^{m}:{15.4}^{s}$  & ${61}^{\circ}:{38'}:{53.0''}$  \\
W3 & ${2}^{h}:{21}^{m}:{42.5}^{s}$  & ${61}^{\circ}:{52'}:{22.0''}$  \\ 
S255 & ${6}^{h}:{09}^{m}:{58.4}^{s}$  & ${18}^{\circ}:{00'}:{19.0''}$  \\
S235B & ${5}^{h}:{37}^{m}:{31.8}^{s}$  & ${35}^{\circ}:{40'}:{18.0''}$  \\
S106 & ${20}^{h}:{25}^{m}:{39.8}^{s}$  & ${37}^{\circ}:{12'}:{46.0''}$  \\
Serpens & ${18}^{h}:{27}^{m}:{25.0}^{s}$  & ${01}^{\circ}:{12'}:{00.0''}$  \\
DR21 & ${20}^{h}:{37}^{m}:{14.0}^{s}$  & ${42}^{\circ}:{09'}:{00.0''}$  \\
Mon R2 & ${6}^{h}:{05}^{m}:{14.0}^{s}$  & ${-6}^{\circ}:{23'}:{00.0''}$  \\
NGC2264  &  ${6}^{h}:{38}^{m}:{26.0}^{s}$  & ${9}^{\circ}:{32'}:{00.0''}$ \\ 
OMC-2 & ${5}^{h}:{32}^{m}:{57.0}^{s}$  & ${-5}^{\circ}:{12'}:{12.0''}$  \\ 
$\rho$ Oph A & ${16}^{h}:{23}^{m}:{17.9}^{s}$  & ${-24}^{\circ}:{17'}:{18.0''}$\\
NGC2024 &  ${5}^{h}:{39}^{m}:{12.0}^{s}$  & ${-1}^{\circ}:{56'}:{40.0''}$ \\
\hline
\end{tabular}
\label{tab_sources}
\end{table}

NGC2024 is not part of the SWAS sample but we have observed this region
as a standard for comparison: it is relatively close ($\approx 450$~pc) and
has been studied already by numerous authors using various techniques. An 
extended map of the cloud and its environment in CS 2--1 was
provided by \cite{Lada91}, and line profiles in four transitions
of the main CS isotope are given by Lada et al. (1997). \cite{Mezger}
have identified seven clumps in NGC2024 from dust observations using the
IRAM 30m telescope 
whereas the FCRAO and KOSMA beams can only distinguish between the two bright
clumps FIR3 and FIR5. The data analysis is performed for the position of 
the brightest clump FIR5. Here, the KOSMA beam also
contains weak contributions from FIR4, FIR6, and FIR7.

\begin{table*}
\caption{Measured line parameters at the central core positions}
\begin{tabular}{lrrr@{\hspace{0.5cm}}rrr@{\hspace{0.5cm}}rrr}
\hline
& \multicolumn{3}{c}{CS 2--1}
& \multicolumn{3}{c}{CS 5--4} 
& \multicolumn{3}{c}{CS 7--6} \\
\cline{2-4} 
\cline{5-7} 
\cline{8-10} 
\multicolumn{1}{c}{Source}
& \multicolumn{1}{c}{$T\sub{mb}$}
& \multicolumn{1}{c}{$\int T\sub{mb} dv$}
& \multicolumn{1}{c}{$\Delta v$}
& \multicolumn{1}{c}{$T\sub{mb}$}
& \multicolumn{1}{c}{$\int T\sub{mb} dv$}
& \multicolumn{1}{c}{$\Delta v$}
& \multicolumn{1}{c}{$T\sub{mb}$}
& \multicolumn{1}{c}{$\int T\sub{mb} dv$}
& \multicolumn{1}{c}{$\Delta v$} \\
& \multicolumn{1}{c}{[K]}
& \multicolumn{1}{c}{[K km/s]}
& \multicolumn{1}{c}{[km/s]}
& \multicolumn{1}{c}{[K]}
& \multicolumn{1}{c}{[K km/s]}
& \multicolumn{1}{c}{[km/s]}
& \multicolumn{1}{c}{[K]}
& \multicolumn{1}{c}{[K km/s]}
& \multicolumn{1}{c}{[km/s]} \\
\hline
W49A$^{(a)}$     &  3.77  &  35.17  &  8.75 $\pm$ 0.55  & 1.54  & 12.61  &  7.69 $\pm$ 0.19& 1.23  & 12.15  &  9.28 $\pm$ 1.24\\
W49A$^{(b)}$     &  4.34  &  32.76  &  7.10 $\pm$ 0.38  & 2.04  & 18.70  &  8.62 $\pm$ 0.15& 1.06  & 9.52   &  8.43 $\pm$ 1.03\\
W33             &  9.74  &  67.24  &  6.49 $\pm$ 0.16  & 3.76  & 27.40  &  6.84 $\pm$ 0.05& 2.94  & 17.44  &  5.57 $\pm$ 0.05\\
W51A            &  10.86 &  110.51 &  9.56 $\pm$ 0.09  & 4.99  & 67.89  &  12.78$\pm$ 0.09& 3.38  & 40.63  & 11.28 $\pm$ 0.08\\
W3(OH)          &  5.50  &  26.48  &  4.53 $\pm$ 0.20  & 2.13  & 10.13  &  4.46 $\pm$ 0.08& 1.25  & 5.81   &  4.40 $\pm$ 0.31\\
W3              &  7.67  &  38.97  &  4.77 $\pm$ 0.05  & 1.35  &  7.72  &  5.36 $\pm$ 0.28& 1.56  & 8.44   &  5.08 $\pm$ 0.09\\
S255            &  8.02  &  21.55  &  2.53 $\pm$ 0.04  & 2.91  & 10.13  &  3.28 $\pm$ 0.03& 1.67  & 5.29   &  2.99 $\pm$ 0.15\\
S235B           &  6.38  &  16.21  &  2.39 $\pm$ 0.05  & 0.85  &  2.67  &  2.97 $\pm$ 0.13& 0.42  & 1.10   &  2.53 $\pm$ 0.19\\
S106            &  3.29  &   8.19  &  2.34 $\pm$ 0.10  & 0.83  &  2.11  &  2.40 $\pm$ 0.22& 0.65  & 2.23   &  3.23 $\pm$ 0.78\\
Serpens         &  3.74  &   9.36  &  2.35 $\pm$ 0.13  & 0.67  &  2.52  &  3.59 $\pm$ 0.35& 0.39  & 1.68   &  4.11 $\pm$ 0.65 \\
DR21            &  6.48  &  25.00  &  3.62 $\pm$ 0.10  & 2.41  &  9.74  &  3.80 $\pm$ 0.18& 2.92  & 10.92  &  3.51 $\pm$ 0.06\\
Mon R2          &  5.53  &  12.26  &  1.90 $\pm$ 0.16  & 2.15  &  6.52  &  2.38 $\pm$ 0.27& 1.70  & 3.83   &  1.95 $\pm$ 0.19\\
NGC2264         &  5.64  &  21.90  &  3.66 $\pm$ 0.08  & 2.30  &  9.48  &  3.89 $\pm$ 0.08& 1.25  & 5.42   &  3.91 $\pm$ 0.12\\
OMC-2           &  5.62  &   9.26  &  1.55 $\pm$ 0.05  & 1.96  &  5.93  &  2.85 $\pm$ 0.10& 0.92  & 1.54   &  1.58 $\pm$ 0.25\\
$\rho$ Oph A    &  4.22  &  11.10  &  2.47 $\pm$ 0.25  & 2.07  &  2.30  &  1.04 $\pm$ 0.08& $<$0.2    &    -    &    -  \\
\hline
\end{tabular}
\label{tab_lineparams}
\end{table*}

\subsection{Observational details}

All sources except NGC2024 were observed in CS 2--1 by Howe ({\it priv.
comm.}) using the FCRAO 14m telescope providing a resolution  of 53$''$.
and a main beam efficiency $\eta\sub{mb}^{2-1}=0.58$. The cores were 
covered by 30-point maps with a sampling of 50$''$.
The CS 2--1 spectra for the southern core in NGC2024 were taken
from Lada et al. (1997).

The CS 5--4 and 7--6 observations used the dual channel KOSMA SIS receiver
with noise temperatures of about 95~K in the 230~GHz branch and 120~K
in the 345~GHz branch. The 3m telescope provides a spatial resolution 
of 110$''$ in CS 5--4 and 80$''$ in CS 7--6.  The default observing mode
for all cores were cross scans with a separation of 50$''$ between 
subsequent points.  Only for NGC2024 complete 5$'\times$5$'$ maps were 
obtained. At the time of the observations the telescope surface 
provided main beam efficiencies $\eta\sub{mb}^{5-4}=0.54$ and 
$\eta\sub{mb}^{7-6}=0.48$ respectively for the two transitions. 
For all measurements we use the conservative estimate of about 10\% 
uncertainty for the main beam efficiency, another 10\% atmospheric 
calibration uncertainty and add another 5\% for possible drifts etc. 
As systematic errors they might sum up linearly to a total
calibration error of at most 25\%.

The FCRAO spectrometer had a channel width of 19.5~kHz
corresponding to a velocity spacing of 0.060~km/s. For the broad
lines from the sample four velocity channels were binned.
The resulting r.m.s. falls between 0.2 and 0.45~K.
The KOSMA spectra were taken with the medium resolution spectrometer
(MRS) and the low resolution spectrometer (LRS) providing channel widths
of 167 kHz and 688 kHz, respectively. Depending on the 
different combinations of these backends with the receivers
at 245 and 343 GHz we obtain velocity spacings between 0.15 and 
0.84 km/s. The particular spacing is not important for the analysis 
performed here because none of the lines shows strong spectral
substructure.
All points were integrated up to a noise limit of 0.1~K
per channel. 

\subsection{Results}

For most sources the CS 2--1 maps show an approximately elliptical 
intensity peak with a weak elongation at scales of a few times the 
resolution. In W49A, Serpens, DR21, Mon R2, OMC-2, $\rho$ Oph A,
and NGC2024 we can distinguish a second intensity
maximum apart from the central position. Tab. \ref{tab_lineparams}
summarises the parameters of the line profiles at the central
position for all cores. The majority of line profiles are approximately
Gaussian as indicated by integrated line intensities close
to the Gaussian value of $1.06\,T\sub{mb}\Delta v$ in Tab.
\ref{tab_lineparams}. Broad wings are only visible in S255, W33, and DR21.

W49 shows a double-peak structure  which has been 
interpreted e.g. by \cite{Dickel} as the footprint of large scale 
collapse. They fitted HCO$^{+}$ line profiles by a spherical 
collapse model but concluded that additional components are 
needed to explain the observations. Using the radiative transfer code
from Appx.\ B we have tested their infall model and found 
that, while reproducing the HCO$^{+}$ profiles, 
it completely fails to explain the CS observations. 
The enhanced blue emission characteristic for collapse is 
visible only in CS 7--6. In CS 2--1 and 5--4 we rather find an enhanced
red emission. No spherically symmetric collapse model
can explain these observations. Instead of constructing
a more complex model we have simply decomposed the
emission into two separate components with a relative
velocity of 8.5~km/s in the line of sight denoted as W49A$^{\rm (a)}$ 
and W49A$^{\rm (b)}$. From the modelling in Sect. 4 it turns 
out that we cannot even distinguish whether the two components are
moving towards each other or apart as long as they do not line up
exactly along the line of sight. Hence, we will treat them
separately in the following, ignoring any possible interaction.
Further observations including other tracers should be included
to better resolve the situation.

\begin{table}
\caption{Measured line parameters within NGC2024}
\begin{tabular}{llll}

\hline
\multicolumn{1}{c}{Transition}
& \multicolumn{1}{c}{$T\sub{mb}$}
& \multicolumn{1}{c}{$\int T\sub{mb} dv$} 
& \multicolumn{1}{c}{$\Delta v$} \\
& \multicolumn{1}{c}{[K]}
& \multicolumn{1}{c}{[K km/s]}
& \multicolumn{1}{c}{[km/s]} \\
\hline
C$^{32}$S 5--4  & 5.19  &   11.65 &  2.10 $\pm$ 0.02 \\ 
C$^{32}$S 7--6  & 3.63  &   9.13 &  2.38 $\pm$ 0.08 \\ 
C$^{34}$S 5--4  & 1.19  &   2.22 &  1.75 $\pm$ 0.09 \\ 
C$^{34}$S 7--6  & 0.65  &   0.98 &  1.45 $\pm$ 0.13 \\ 
\hline
\end{tabular}
\label{tab_ngc2024params}
\end{table}

In NGC2024 we also mapped the less abundant isotope 
C$^{34}$S with KOSMA in addition to the main CS isotope
observed in all cores. The observed line 
parameters are given in Tab. \ref{tab_ngc2024params}.
Lada et al. (1997) provided detailed CS spectra for the 
southern core at the position of the FIR5.
They obtained in the CS 2--1 transition
$T\sub{mb}=15.4$~K, $\Delta v=1.80$~km/s at 24$''$ resolution,
in CS 5--4  $T\sub{mb}=9.6$~K, $\Delta v=2.1$~km/s at 30$''$
resolution, in CS 7--6 $T\sub{mb}=10.1$~K, $\Delta v=2.7$~km/s
at 20$''$ resolution, and in CS 10--9 $T\sub{mb}=10.6$~K,
$\Delta v=2.1$~km/s at 14$''$ resolution.

\section{Cloud parameters from the escape probability model}
\label{sect_ep}

To interpret the small amount of information contained in
observations of at most five transitions showing
mainly Gaussian profiles and essentially unresolved
approximately circular symmetric intensity distributions,
we need a simple cloud model
that is both physically reasonable and characterised
by few parameters.  An obvious choice is a spherically symmetric
model. This geometry reflects early phases and the large scale 
behaviour of several collapse simulations (e.g.
\cite{Galli99}), whereas the inner parts of collapsing clouds
are probably flattened structures
(e.g. \cite{Li}). 

Even in spherical geometry there is no simple way to
solve the radiative transfer problem relating the
cloud parameters to the emitted line intensities (see Appx.\ A.1). 
Thus we cannot compute the cloud properties directly from the
observations.

\subsection{Application of the escape probability approximation}

A common approach is the escape probability 
approximation discussed in detail in Appx.\ A.2. Assuming that all 
cloud parameters, including the excitation temperatures, are constant
within a spherical cloud volume one can derive a simple
formalism relating the three parameters kinetic temperature 
$T\sub{kin}$, gas density $n\sub{H_2}$, and column density 
of radiating molecules on the scale of the global velocity 
variation $N\sub{mol}/\Delta v$ to 
the line intensity at the cloud model surface.
No assumption on molecular abundances is required.

Since a telescope does not provide a simple pencil beam
we have to correct the model surface brightness 
temperature by the beam filling factor $\eta\sub{f}$, given as 
the convolution integral of the normalised intensity distribution with
the telescope beam pattern, to compute the observable beam temperature.
Unfortunately, the brightness profile of the source is a non-analytic function
where we can only give simple expressions for the central value
observed in a beam much smaller than the source or 
for the integral value observed in a beam much larger than the source
(Eqs. \ref{eq_tmb_lvg} and \ref{eq_tmb_esc}).
For intermediate situations we approximate the beam temperature
by starting from both limits and using a beam filling factor given by
the convolution integral of two Gaussians.
The difference between the two values provides an estimate of 
the error made in the beam convolution.

\begin{table}
\caption{Source size corrected for beam convolution.}
\begin{tabular}[h]{lllllll}
\hline
\multicolumn{1}{c}{Source}
& \multicolumn{3}{c}{ FWHM in $\alpha$ $[']$}
& \multicolumn{3}{c}{ FWHM in $\delta$ $[']$}
\\ \hline
\multicolumn{1}{c}{}
& \multicolumn{1}{l}{ 2 -- 1 }
& \multicolumn{1}{l}{ 5 -- 4 }
& \multicolumn{1}{l}{ 7 -- 6 }
& \multicolumn{1}{l}{ 2 -- 1 }
& \multicolumn{1}{l}{ 5 -- 4 }
& \multicolumn{1}{l}{ 7 -- 6 }
\\ \hline
     W49A &  1.0  &  1.1  &  0.3  &  1.5  &  1.7  &  1.3  \\
     W33 &  1.7  &  1.7  &  1.4  &  1.7  &  1.3  &  1.0  \\
    W51A &  1.9  &  1.7  &  1.7  &  2.1  &  2.2  &  1.1  \\
  W3(OH) &  2.1  &  1.1  &  1.3  &  1.7  &  1.4  &  1.4  \\
      W3 &  1.0  &  0.6  &  0.8  &  1.0  &  0.6  &  0.3  \\
    S255 &  1.9  &  1.6  &  1.1  &  1.2  &  0.6  &  1.1  \\
   S235B &  1.0  &  2.2  &  1.1  &  0.9  &  0.2  &  0.3  \\
    S106 &  3.0  &  1.6  &       &  2.5  &  2.2  &       \\
    Serpens & 2.0 & 2.2 && 2.1 & 2.3  \\
    DR21 &  3.0  &  2.6  &  1.8  &  1.5  &  1.7  &  1.8  \\
  Mon R2 &  3.2  &  3.0  &  1.1  &  3.6  &  3.2  &  1.6  \\
 NGC2264 &  3.2  &  1.7  &  0.6  &  2.9  &  2.1  & \\
   OMC-2 &  2.4  &  1.4  &  0.6  &  2.5  &  1.4  &  0.6  \\
    $\rho$ Oph A &  1.3  &  3.0  &  2.6  &  1.9 & 2.9 & 2.5 \\
 NGC2024 &  1.1  &  1.1  &  1.3  &  1.3  & 2.5   & 2.3 \\
\hline
\end{tabular}
\label{tab_coresize}
\end{table}

To compute the integral we fitted the observed brightness distributions
by Gaussians. Most cores are well approximated 
by slightly elongated Gaussians. W49A, S235B, Serpens, and $\rho$ Oph A
show asymmetric scans so that the size determination is somewhat 
uncertain. {\changed The fit error is about 0.3$'$ for these three sources.
For the rest of the cores we obtain typical values of less than 0.2$'$.}
The true object size finally follows from the deconvolution
of the measured intensity distribution with the telescope beam.
The resulting source sizes in $\alpha$ and $\delta$ are  given in
Tab. \ref{tab_coresize}. As the geometric mean is sufficient to
compute the beam filling factor we don't expect any serious error
from the fact that the cross-scans in $\alpha$ and $\delta$ do
not necessarily trace the major axes of the brightness distribution.
For sources which are considerably smaller than the
beam widths of 53$''$, 107$''$, and 80$''$, respectively,
only a rough size estimate is possible
according to the nonlinearity of the deconvolution. This holds
for W49A, W3, S235B, and partially S255. Most clouds, however, show
an extent of the emission which is close to the beam size.

In general different values are obtained for the spatial FWHMs
in the different lines. In Tab. \ref{tab_coresize} we find two 
classes of sources
with respect to the variation of the source size depending on the
transition observed. Most cores show a monotonic decrease of the 
visible size when going to higher transitions. This is expected 
from the picture that higher transitions are only excited in
denser and smaller regions. Serpens, $\rho$ Oph A, and NGC2024, however
show the smallest width of the fit in the CS 2--1 transition. This 
is explained by eye inspecting the 2--1 maps and corresponding 
high-resolution observations from the literature where we see that the three 
sources break up into several clumps which are only separated in the
53$''$ beam but unresolved in the KOSMA beams.
In these cases, we have restricted the analysis to the major core 
seen in the CS 2--1 maps using its size to compute the beam filling,
although this approach introduces a small error in the data analysis
by assigning the whole flux measured in the higher transitions
to this central core.

Applying the two limits for the beam size treatment 
(Eqs. \ref{eq_tmb_lvg} and \ref{eq_tmb_ep}) using the sizes
from Tab. \ref{tab_coresize} we find that the resulting gas and 
column densities are the same within 20\% except for W49A, W3, 
S235B, and NGC2264.
The first three are small compared to the beams so that 
the results from Eq. (\ref{eq_tmb_lvg}) have to be rejected and
only Eq. (\ref{eq_tmb_ep}) can be used.
For NGC2264 we cannot provide a simple explanation for the difference
so that we give a relatively large error bar covering the results
from both approximations.

The size of the resulting parameter range in $T\sub{kin}$, $n\sub{H_2}$, 
and $N\sub{CS}/\Delta v$ is determined by the accuracy of the
observations. For two cores it was only possible to set a lower limit 
to the gas density.
Moreover, we were not able to provide any good constraint to the cloud 
temperature for all sources. Values between about 30~K and 150~K are 
possible. 
Hence, an independent determination of the cloud temperatures is required.
Several different methods based on optically thick CO, NH$_3$ or
dust observations are discussed in the literature and we used the
values from the references given in Tab. \ref{tab_epparm}. In addition
to these values we also used 50 K as assumed by \cite{Plume} as
``standard'' temperature in the parameter determination for 
massive cores.

\subsection{Resulting core parameters}

\begin{table}
\caption{Clump parameters derived from the escape probability model}
\begin{tabular}{llrlrl}
\hline
\multicolumn{1}{c}{Source}
& \multicolumn{1}{c}{ $T_{kin}$}
& \multicolumn{1}{c}{ $n\sub{H_{2}}$} 
& \multicolumn{1}{c}{ $\!\!\times/\div$} 
& \multicolumn{1}{c}{ $N\sub{CS}/\Delta v$ }
& \multicolumn{1}{c}{ $\!\!\times/\div$} 
\\ 
\multicolumn{1}{c}{ }
& \multicolumn{1}{c}{ [K] }
& \multicolumn{1}{c}{ [cm$^{-3}$] }
& & \multicolumn{1}{c}{ [cm$^{-2}$/kms$^{-1}$] } 
\\ \hline
W49A$^{(a)}$    & 20$^{\mathrm{a}}$ & $\!\!>$ 7.3\,10$^6$&        & 2.2\,10$^{14}$ & 1.3\\
        & 50$^{\mathrm{a}}$ & 1.3\,10$^6$ & 1.5 & 2.1\,10$^{14}$ & 1.2\\
W49A$^{(b)}$    & 20$^{\mathrm{a}}$ & 4.5\,10$^6$& 1.6 & 2.8\,10$^{14}$ & 1.3\\
        & 50$^{\mathrm{a}}$ & 7.6\,10$^5$ & 1.5 & 2.7\,10$^{14}$ & 1.2\\
W33     & 40$^{\mathrm{b}}$ & 1.7\,10$^6$ & 1.7 & 2.3\,10$^{14}$ & 1.2\\
        & 50$^{\mathrm{c}}$ & 1.2\,10$^6$ & 1.6 & 2.3\,10$^{14}$ & 1.2\\
W51A    & 20$^{\mathrm{a}}$ & 2.1\,10$^7$ & 1.6 & 4.8\,10$^{14}$ & 1.7\\
        & 50 & 1.1\,10$^6$ & 1.5 & 4.1\,10$^{14}$ & 1.4\\
        & 57$^{\mathrm{a}}$ & 8.9\,10$^5$ & 1.4 & 3.7\,10$^{14}$ & 1.4\\
W3(OH)  & 30$^{\mathrm{d}}$ & 1.9\,10$^6$ & 1.5 & 1.3\,10$^{14}$ & 1.2\\ 
        & 50 & 8.5\,10$^5$ & 1.4 & 1.3\,10$^{14}$ & 1.2\\
W3      & 30$^{\mathrm{e}}$ & 7.1\,10$^6$ & 2.9 & 7.4\,10$^{14}$ & 1.5\\
        & 50 & 9.8\,10$^5$ & 1.8 & 8.1\,10$^{14}$ & 1.4\\
        & 55$^{\mathrm{f}}$ & 7.4\,10$^5$ & 1.6 & 8.3\,10$^{14}$ & 1.3\\
S255    & 40$^{\mathrm{g}}$ & 1.3\,10$^6$ & 1.6 & 1.3\,10$^{14}$ & 1.3\\
        & 50 & 9.3\,10$^5$ & 1.5 & 1.2\,10$^{14}$ & 1.3\\
S235B   & 40$^{\mathrm{h}}$ & 1.2\,10$^6$ & 1.5 & 3.7\,10$^{13}$ & 1.3\\
        & 50 & 8.9\,10$^5$ & 1.5 & 3.6\,10$^{13}$ & 1.3\\
S106    & 10$^{\mathrm{i}}$ & $\!\!>$6.3\,10$^6$ &        & 4.7\,10$^{13}$ & 1.4\\
        & 25$^{\mathrm{i}}$ & 7.1\,10$^5$ & 1.5 & 3.8\,10$^{13}$ & 1.4\\
        & 50 & 2.5\,10$^5$ & 1.6 & 3.9\,10$^{13}$ & 1.4 \\
Serpens & 25$^{\mathrm{j}}$ & 4.6\,10$^5$ & 1.2 & 2.8\,10$^{13}$ & 1.5\\
        & 50 & 1.9\,10$^5$ & 1.6 & 2.8\,10$^{13}$ & 1.5 \\
DR21    & 35$^{\mathrm{k}}$ & 1.5\,10$^6$ & 1.7 & 9.3\,10$^{13}$ & 1.9\\
        & 50 & 9.1\,10$^5$ & 1.6 & 8.7\,10$^{13}$ & 1.4\\
Mon R2  & 25$^{\mathrm{l}}$ & 1.7\,10$^6$ & 1.8 & 4.5\,10$^{13}$ & 1.5\\
        & 50$^{\mathrm{m}}$ & 6.2\,10$^5$ & 1.7 & 3.7\,10$^{13}$ & 1.7\\
NGC2264 & 25$^{\mathrm{n}}$ & 1.4\,10$^6$ & 2.4 & 7.9\,10$^{13}$ & 2.6\\
        & 50 & 4.7\,10$^5$ & 2.3 & 7.8\,10$^{13}$ & 2.5\\
OMC-2   & 19$^{\mathrm{q}}$ & 1.9\,10$^6$ & 1.6 & 6.3\,10$^{13}$ & 1.4\\
        & 24$^{\mathrm{o,p}}$ & 1.1\,10$^6$ & 1.5 & 6.2\,10$^{13}$ & 1.4\\
        & 50 & 3.4\,10$^5$ & 1.7 & 6.2\,10$^{13}$ & 1.4\\
$\rho$ Oph A  & 25$^{\mathrm{r}}$ & 9.5\,10$^5$ & 1.6 & 2.0\,10$^{13}$ & 1.4\\
        & 50 & 3.7\,10$^5$ & 1.8 & 2.0\,10$^{13}$ & 1.5 \\
NGC2024 & 25$^{\mathrm{s}}$ & 9.1\,10$^6$ & 2.9 & $>$4.5\,10$^{14}$\\
        & 40$^{\mathrm{t}}$ & 2.1\,10$^6$ & 2.0 & $>$3.5\,10$^{14}$\\
        & 50  & 1.6\,10$^6$ & 1.7 & $>$3.5\,10$^{14}$\\
\hline
\label{tab_epparm}
\end{tabular}
\\
\footnotesize
\mbox{}$^{\mathrm{a}}$ \cite{Sievers} ,
$^{\mathrm{b}}$ \cite{Goldsmith}, \\
$^{\mathrm{c}}$ \cite{Haschick}, 
$^{\mathrm{d}}$ \cite{Wilson},  \\
$^{\mathrm{e}}$ \cite{Tieftrunk98},
$^{\mathrm{f}}$ \cite{Tieftrunk95},\\
$^{\mathrm{g}}$ \cite{Jaffe}, 
$^{\mathrm{h}}$ \cite{Nakano}, \\
$^{\mathrm{i}}$ \cite{Roberts97},
$^{\mathrm{j}}$ \cite{McMullin},  \\
$^{\mathrm{k}}$ \cite{Garden},
$^{\mathrm{l}}$ \cite{Montalban}, \\
$^{\mathrm{m}}$ \cite{Giannakopoulou},
$^{\mathrm{n}}$ \cite{Kruegel},  \\
$^{\mathrm{o}}$ \cite{Castets},
$^{\mathrm{p}}$ \cite{Batrla}, \\
$^{\mathrm{q}}$ \cite{Cesaroni}, 
$^{\mathrm{r}}$ \cite{Liseau}, \\
$^{\mathrm{s}}$ \cite{Ho},
$^{\mathrm{t}}$ \cite{Mezger}
\normalsize
\end{table}

Tab. \ref{tab_epparm} lists the parameters from the escape 
probability model for all
cores. Whereas the column density is well constrained for
most clouds, there is a considerable uncertainty in the gas 
density resulting from the unknown cloud temperature. 
At the temperature of 50 K we obtain average values and logarithmic 
standard deviation factors of
\begin{eqnarray}
\langle n\sub{H_{2}}\rangle =& 7.9\,10^5 \qquad &\times\!/\!\div 1.5 \\
\langle N\sub{CS}/\Delta v\rangle =& 1.2\,10^{14} \qquad &\times\!/\!\div 2.8\nonumber
\end{eqnarray}

From the 71 cores analysed by \cite{Plume} assuming this temperature
they obtained 
\begin{eqnarray}
\langle n\sub{H_{2}}\rangle =& 8.5\,10^5 \qquad &\times\!/\!\div 1.7 \\
\langle N\sub{CS}/\Delta v\rangle =& 2.5\,10^{14} \qquad &\times\!/\!\div 3.1\nonumber
\end{eqnarray}

This good agreement indicates first that both investigations study 
the same type of clouds visible in the CS transitions. Second, this 
shows that the static escape probability (Eq. \ref{eq_tmb_statep}) used here 
and the LVG escape probability (Eq. \ref{eq_tmb_lvg}) applied by \cite{Plume} 
differ only marginally as discussed already by \cite{Stutzki85}. 
Third, the observational data from both telescopes give
almost equivalent results, i.e. the reliability of the parameters
hardly profits from using the 10 times smaller beam of the IRAM telescope 
when the escape probability approximation is used.

For NGC2024 we are able to test the consistency of the results
from the CS and the C$^{34}$S observations. The resulting hydrogen 
densities agree for both isotopes within 20\% at all temperatures 
assumed. We obtain 9\,10$^6$ cm$^{-3}$ at 25 K (Mezger et al. 1992) and
3.2\,10$^6$ cm$^{-3}$ at 40 K (Ho et al. 1993). Unfortunately, this
is a core where we can only give a lower limit to the column densities. 
The limits deviate by a factor 13, which is significantly
different from the terrestrial isotopic ratio of 23 but close
to the value 10 derived by \cite{Mundy} for the isotopic ratio in NGC2024. 

The escape probability model provides a first estimate
to the physical parameters but its limitations are obvious.
It is definitely not justified to assume constant parameters
within the whole cloud. Moreover, several observations are in contradiction to
the parameters from the escape probability models.
\cite{Lada97} detected the CS 10--9 transition in NGC2024 and
\cite{Plume} observed the 10--9 and 14--13 transitions in
S255 and W3(OH). The critical densities for these transitions are
about 6\,10$^7$ cm$^{-3}$ and 2\,10$^8$ cm$^{-3}$ respectively.
From the densities in Tab. \ref{tab_epparm} 
one would conclude that these transitions 
are not excited. Hence, a more sophisticated model has to be applied 
to obtain a physically reasonable explanation of the measurements.
\cite{Plume} suggested a two-component model or continuous
density gradients to resolve this contradiction. We will discuss a 
self-consistent radiative transfer model including a radial
density profile in the following.
\label{sect_epresults}

\section{Cloud parameters from the nonlocal model}
\label{sect_nonlocal}

\subsection{Line fitting by {\it SimLine}}

We performed nonlocal radiative transfer simulations using
the line radiative transfer code {\it SimLine} introduced in detail
in Appx.\ B.
{\it SimLine} is a FORTRAN code to compute the profiles of
molecular rotational lines in spherically symmetric clouds with 
arbitrary density, temperature and velocity distribution.
It consists of two parts:
the self-consistent solution of the balance equations for all level
populations and energy densities at all radial points
and the computation of the emergent line profiles observed
by a telescope with finite beam width and arbitrary offset. 
The optical depths in the lines may vary from minus a few, corresponding
to weak masing, to several thousand.

Already in the spherically symmetric description of a core 
we face a large number of parameters. For all quantities
(hydrogen density, kinetic temperature, velocity dispersion,
molecular abundances)
a radial function has to be found. Regarding the limited
amount of information available from the three transitions,
this leaves many options open. We decided to assume
simple power-law radial functions and a central region with constant
parameters for all quantities in the core simulations. This reduces
the number of parameters to two (central value and radial exponent) 
for each gas property, plus the outer and inner radius. 

The parameter fit procedure used the multidimensional downhill simplex
algorithm from Press et al. (1992). Although it is not
the most efficient way in terms of convergence speed it turned
out to be very robust in all situations considered. Because a downhill
simplex code does not necessarily find the global minimum of
the $\chi^2$ function we performed for each core several runs
with randomly chosen initial simplex covering a large part of the
physically reasonable parameter space. For all clouds we made at
least 30 runs to get a rough idea of the topology of the $\chi^2$
function. For cores like W33 this turned out to be sufficient
since only one large minimum showed up which was found in almost
half of the runs. The other extreme is S106 where we needed
almost 1000 runs to be sure that we found the global minimum.
Here, the $\chi^2$ function was quite complex with numerous
local minima. Future improvements of the fit procedure should include 
more sophisticated algorithms like simulated-annealing approaches.

The noise in the line profiles produces some graininess of the $\chi^2$
function when directly fitting the measured profiles. This
results in a very slow convergence of the algorithm close to the
minimum. A considerable acceleration can be
obtained by not fitting the measured noisy data but a smooth approximation
to them. The line profiles were represented by a superposition of
a Gaussian and a Lorentzian profile. This allows a good characterisation
of all measured profiles including the reproduction of asymmetric
profiles, self-absorption dips and line wings. 

Taking the three measured transitions and their
spatial extent as the quantities to be reproduced by a $\chi^2$ fit
we find that we are able to derive at most six parameters to
a reasonable accuracy. Fits with seven or eight free parameters,
although still slightly improving the numerical $\chi^2$ value, do
not produce any significant changes above the noise limit. 
This is comparable to the results by \cite{Young} fitting the full
position velocity map of a particular core in a single transition.
There is however a number of additional parameters
where the model can derive certain limits.
Hence, we have to decide first which parameters should be fitted 
directly, which parameters may be constrained by preventing
reasonable fits when outside a certain range, and which parameters
can be guessed independently from a physical line of reasoning. This is
discussed in detail for all quantities in the following subsections.

Unfortunately,
it is impossible to give an comprehensive error estimate for
the parameters derived from the simulations. This would need
a description of the six-dimensional surface in the parameter space
within which none of the observational error bars is exceeded.
According to the complex topology of the $\chi^2$ function
it is not possible to give an easy description for the six-dimensional
valley around the global minimum or its boundaries.

As a simple alternative we performed only one-dimensional 
variations to get a rough estimate
of the maximum error in the parameters that we must expect. After
the convergence of the $\chi^2$ fit, each of the fit parameters 
was varied up and down until one of the computed lines
deviated by the assumed maximum observational error of 25\,\%
from the measured lines. The central values of the different
functions were varied independently. When changing the
inner radius, the central values of the gas parameters were adjusted
to keep the functions in the power-law region unchanged.
When varying exponents the corresponding central values were 
corrected in such a way that the parameters at the density of 
2\,10$^{6}$cm$^{-3}$ remained constant. By the selection of this 
density as the fix point when changing the slope, we scan about the
maximum possible range of the exponents. This provides a 
conservative estimate of the maximum error.

\subsection{The turbulence description}

In spite of the large number of free parameters in the cloud models
it is impossible to fit the line profiles with a smooth density and
velocity distribution. The typical self-absorbed profiles known for 
all microturbulent codes (\cite{White}) appear in this case.
Thus we have to take into account the effects of internal clumping
and turbulence leading to a more realistic picture and preventing
strong self-absorption.

{\it SimLine} treats turbulence and clumping in a local statistical 
approximation following \cite{MSH}
(see Appx.\ B.2). The cloud material is subdivided into small 
coherent units (clumps) with only thermal internal velocity dispersion. 
The relative motion of many units then provides the full velocity 
profile. \cite{MSH} showed that the effective optical depth of 
such an ensemble can be considerably reduced 
compared to the microturbulent approximation. The relative reduction
of the total optical depth depends on the optical depth, i.e. of size
the coherent units. 

This description does not need any 
assumptions on the nature of the turbulence creating the internal
cloud structure. The reduction occurs in the same way whether
clumps are units of the same velocity in a medium
of constant density, representing the behaviour of incompressible turbulence,
or whether they are density enhancements in a thin inter-clump medium.
The different nature of these two scenarios has to be taken into account,
however, when computing the excitation.
Both the average gas density, providing a measure for the column density
and thus the line intensity, and the local density, providing the
collisional coupling to the gas, enter the balance equations. In case
of coherent units in velocity space both densities agree.

For density clumps we use the additional simplification to treat the
cloud locally as a two-component medium neglecting the contribution of
the inter-clump medium to the radiative transfer. Then the collisional 
excitation is provided by the density in the clumps and the column 
density is provided by the average density.  The ratio between the two 
quantities reflects the filling factor of the volume occupied by 
dense clumps. 
From the mathematical point of view this is equivalent to the
treatment of the cloud as a homogeneous medium where the
clumping occurs only in velocity space and the abundance of the
radiating molecules is reduced. Hence, it is impossible to separate
the influence of the filling factor from that of the molecular abundance
so that the line fit provides only a combined quantity
which we will denote combined abundance in the following.
Beside the modification of the molecular abundance the statistical 
turbulence description introduces the size of the coherent units as 
an additional parameter.

\begin{figure}
\epsfig{file=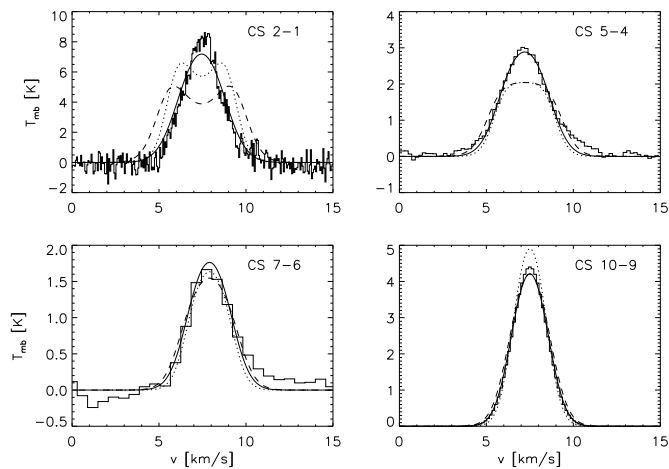, angle=90, width=8.7cm}
\caption{Line profiles observed at the central position of
S255 in four CS transitions together with best-fit models. The
dashed line represents the microturbulent approximation,
the dotted line stands for clumping in velocity space
with cell sizes of 0.01 pc and the solid line includes
the additional effect of density clumping where the combined
abundance is reduced by a factor 100.}
\label{fig_s255_turb}
\end{figure}

Fig. \ref{fig_s255_turb} demonstrates the influence
of the turbulence.  It shows the central line profiles of the
best fits to the S255 observations using different
assumptions on the turbulent nature.
We selected S255 here because we can exploit the
advantage of additional data for the CS 10--9 transition measured by
\cite{Plume} providing more constraints on the cloud 
model\footnote{Unfortunately, \cite{Plume} listed only the peak 
intensity and FWHM so that we have to assume a Gaussian for the 
CS 10--9 profile here.}.
In all models the core is optically thin in CS 7--6 and 10--9,
so that the reduction of the optical depth by the turbulent
clumping produces only minor changes in these lines. In CS 2--1
the differences are, however, most obvious. In the microturbulent
description and for the incompressible turbulence we find
self-absorbed line profiles. A reasonable fit to the
observations is only possible using the turbulence model including
clumps with enhanced density. The reduction of the effective optical 
depth of the cloud by increasing the optical depth of the coherent 
units produces narrower lines providing a better
fit to the observed line profiles. 

We find that the parameter fits do not provide accurate
values for the clump size and the combined abundances reflecting
the volume filling factor. They only constrain an interval
of possible values. We obtained good fits to the observations for 
the full range of clump sizes between about 0.005 and 0.05~pc. 
This size scale is confirmed by independent determinations of clump sizes
by high-resolution observations in some of our cores.
For NGC2024 \cite{Mezger} determined a radius of dense condensations
around 0.015~pc and the interferometric studies by \cite{Wiesemeyer}
showed values between 0.005 and 0.01~pc. In W3 \cite{Tieftrunk98}
observed compact clumps with a size of 0.02~pc and
in OMC-2 \cite{Chini} found dust condensations with radii
between 0.01 and 0.05~pc.

For a better comparison to turbulence theory we prefer
to specify the cell size in terms of the turbulent correlation 
length $l\sub{corr}$ which should be on the order of 0.1~pc 
(Miesch et al. 1994, Goodman et al. 1998). The size of the units 
which are coherent with respect to the line radiative
transfer is smaller by the ratio of the thermal line width to the total
velocity dispersion. Thus the size range
found corresponds to correlation lengths between 0.04 and 0.4~pc. 
In the following computations we use the intermediate value of 0.1~pc as
correlation length for all clouds.

Regarding the combined abundances we find two classes of objects.
The majority of cores, including the example of S255, is well fitted
by values between $10^{-11}$ and 10$^{-10}$, whereas a second class, 
consisting of W3, Serpens, and S106 needs values between 
$10^{-10}$ and 10$^{-9}$ for a good fit. Because we do not know
the molecular abundances there is no way to translate these values 
directly into clump filling factors. Assuming the CS abundances
of $1.3\,10^{-9}$ to $1.3\,10^{-8}$ obtained by \cite{Hatchell}
for several star-forming cores, the first class corresponds to
filling factors around 0.01, whereas the CS abundance of $4\,10^{-10}$
from \cite{Plume} corresponds to a filling factor of 0.1. The filling
factors in the second group are ten times higher accordingly. In all following
computations we have used a combined abundance factor of $3\, 10^{-11}$
for the first and $3\, 10^{-10}$ for the second group. In the translation
to cloud masses we will use the intermediate CS abundance of 
$4\,10^{-9}$.

As an additional parameter quantifying turbulence in a one-dimensional
cloud model we have to take a radial variation of the turbulent 
velocity distribution into account to explain the observed size-line width
and size-density relations (\cite{Larson69}). Exponents of this 
radial dependence between about 0.1 and 0.7 are typically discussed 
(see e.g. \cite{Goodman}).
We left the width and the exponent of the turbulent velocity 
distribution as free parameters and obtained exponents 
between 0.15 and 0.65.
\label{sect_turbulence}

\subsection{The density structure}

Any physically reasonable cloud model should include a spatial dependence
of the gas parameters. {\changed Collapse simulations might provide 
reasonable model assumptions for the density structure.}
\cite{Bodenheimer} and \cite{Shu} have shown that an isothermal sphere 
evolves into a power law density profile $n\sub{H_2}
\propto r^{-\alpha_n}$ with $\alpha_n=2.0$. Homologous collapse
simulations provide an exponent $\alpha_n=3.0$ (Dickel \& Auer 1994)
and the free-fall collapse discussed by \cite{Welch} results in a density 
exponent $\alpha_n=1.5$. The inside-out collapse model (Shu 1997) 
combined two regions of different exponents and recent more 
sophisticated collapse simulations (see e.g. \cite{Basu})
show more complex density structures with an average 
exponent $\alpha_n$ between 1.5 and 1.7. Dust observations of
the density profile of protostellar cores show evidence both for 
cores with a typical $r^{-2}$ profile and for cores with flat 
density structure and a sharp outer edge (\cite{Andre}). {\changed 
Thus we expect exponents between about 1.5 and 2.0 in our power-law 
density model which is bound by an outer cut-off and a central constant
region.} Although this simple model may not reflect
the whole complexity of the density profile we can hardly
derive any more information from the limited observations
available. {\changed We left the central density and the density exponent
as free parameters to be fitted.} The resulting exponents span the
relatively wide range between 1.1 and 2.2.

In the fit of the radii confining the power-law density profile 
we face two problems. We cannot distinguish changes of the model 
parameters on the smallest size scales where even the highest 
observed transition is thermalised because of the high density.
Thus we can only set an upper limit $R\sub{cent}^{\rm max}$
to the radius of the central region where a transition from the power 
law behaviour in the envelope to constant parameters might occur. 
Only in four clouds -- Serpens, Mon~R2, $\rho$ Oph A, and the 
b-component of W49A -- the strength of the CS 7--6 line
sets an upper limit to the density so that we can derive the inner 
radius directly from the observations. In all parameter fits, we left 
the central radius as a free parameter. After the fit we 
increased the radius until the maximum deviation in one 
of the lines reached 5\,\%. This provides a reasonable upper 
limit to the inner radius in all cases where the line profiles are 
independent of the density structure below this limit and gives 
only small modifications in the four cases where the inner 
radius was already well constrained by the fit.

The outer radius of the cloud is also quite uncertain. We can 
easily provide a value for the extent of gas at densities above about
$3\,10^5$ cm$^{-3}$ based on the spatial extent of the CS 2--1 emission
and the line profiles. But it is not possible to derive a reliable value
for the extent of low density gas. We can only give a lower limit to the
outer radius  and thus the mass of the massive cores considered.
As the density exponents are typically shallower than --2, a majority of
mass could be present beyond this radial limit at low
densities invisible in CS.  We excluded the outer radius from the 
parameter fit using a sufficiently large value for all clouds and 
performed later a separate run reducing the radius to find the minimum 
outer radius $R\sub{cloud}^{\rm min}$ in a way equivalent to the 
maximisation of the inner radius described above.
\label{sect_radfunc}

\subsection{The temperature structure}

\begin{table}
\caption{Resulting cloud parameters for S255 assuming different
exponents for the radial temperature dependence and fitting only the central line profiles.}
\begin{tabular}{lrcrcccc}
\hline
$\alpha_T$& 
\multicolumn{1}{c}{$T\sub{c}$}& 
\multicolumn{1}{c}{$\langle T \rangle$}& 
\multicolumn{1}{c}{$\alpha_n$}&
\multicolumn{1}{c}{$\langle n\sub{H_2} \rangle$}& 
\multicolumn{1}{c}{\hspace*{-1mm}$R\sub{cloud}^{\rm min}$}&
\multicolumn{1}{c}{\hspace*{-1mm}$M\sub{cloud}$}&
\multicolumn{1}{c}{\hspace*{-2mm}FWHM$^{\mathrm a}$}\\ 
& \multicolumn{1}{c}{[K]}& 
\multicolumn{1}{c}{[K]}&&
\multicolumn{1}{c}{[cm$^{-3}$]}&
\multicolumn{1}{c}{[pc]}&
\multicolumn{1}{c}{[M$_{\sun}$]}&
\multicolumn{1}{c}{[$'$]} \\
\hline
\hspace*{0.5mm} 0.0 & 47 & 47 & -1.5 & 3300 & 1.1 & 1200 & 1.8 \\
-0.12& 73 & 52 & -1.3 & 2500 & 1.3 & 1500 & 2.2\\
-0.4 & 220 & 52 & -1.0 & 1000 & 2.5 & 7000 & 3.6\\
\hline
\end{tabular}
\\
${}^{\mathrm a}$ CS 2--1 transition
\label{tab_temp}
\end{table}

The temperature distribution of massive cores
is still a matter of debate (cf. \cite{Garay}).
During early phases of cloud collapse the temperature should remain
constant as long as the core remains optically 
thin. Deviations are to be expected, however, as protostellar
sources are formed in most of our cores, leading to an internal 
heating of the cloud. Moreover in thin outer regions external 
heating can be important. Based on several observational results
\cite{Scoville} set up a spherical cloud model with a warm 
inner region resulting in an temperature profile 
$T\sub{kin} \propto r^{\alpha_T}$ with $\alpha_T\approx -0.4$. 

Hence, we should also derive the core temperature and the
temperature exponent from the radiative transfer model.
In a first run we have
investigated the influence of a temperature gradient in S255 when
fitting only the central line profiles. We compared 
the best fit models to the S255 observations using
either a constant cloud temperature, a temperature decaying with
the exponent -0.4, or the temperature exponent as a free parameter.
The latter case provided a best fitting exponent of -0.12. All 
three fits showed an excellent agreement in terms of the central 
line profiles falling almost exactly on the 
solid curves in Fig. \ref{fig_s255_turb}.
Thus, it is impossible to favour a certain 
exponent from the least squares fit of the line profiles only.

In Tab.~\ref{tab_temp} we see the resulting values for the other 
core parameters. The average temperature given here is computed
as the mass-weighted average up to $R\sub{cloud}^{\rm min}$.
As main difference we find a kind of compensation between temperature 
and density exponent. The sum of both exponents is kept approximately
constant to fulfil the constraints given by the line ratios. 
Temperature and density gradient act in a similar manner, leading to 
higher excitation in regions which are either denser or warmer.
The variation of 
the density gradient, however, changes the extent of the emission
in the CS 2--1 transition (the change is much smaller in the higher 
transitions). Consequently, we can constrain the temperature exponent
when taking the observed size of the source into account.  The beam convolved 
FWHM of the CS 2--1 emission in S255 falls between 1.5$'$ and 2.1$'$ 
clearly excluding models with steep temperature gradients. 

In the fit procedure applied to all cores we have thus
included the fit of the spatial extent. In this way we can constrain
the temperature exponent but we are not able to derive exact values.
In the parameter fit we always start with an isothermal cloud and 
change the temperature gradient in steps of 0.1 until the model 
provides a simultaneous good fit to the central line profiles and 
the spatial FWHM of the emission. It turns out that the observations
of all clouds except W33, W3 and Mon R2 can be fitted by an isothermal
model. This indicates that a large part of the gas mass in the clouds is 
characterised by a uniform kinetic temperature. This isothermal
behaviour, however, does not support the assumption of constant
excitation temperatures like in escape probability model. The
excitation temperatures show local variations which are steeper than 
the density gradient. In the derivation of the kinetic temperature
structure we have to keep in mind, however, that we are not very sensitive
to the exact value of the kinetic temperature gradient. Observations 
of higher transitions like CS 14--13 would be needed to obtain 
reliable values including full error estimates. 

\subsection{Other parameters}

Beside the temperature gradient we can expect the formation
of compact \HII regions in the centre of a star forming core.
Using the parameters of the central \HII region derived
by \cite{Dickel} for W49 ($R=0.2$~pc, $n\sub{e}=2.6\,10^4$ cm$^{-3}$,
$T\sub{e}=10^4$~K) we have compared the resulting CS 
lines when either including or neglecting the \HII region
in the radiative transfer computations 
(see Appx.\ \ref{sect_hii_region}). We find that the influence
of the \HII region is negligible in this example for the four
CS transitions considered here. 

{\changed In general \HII regions have two dominant effects
on the line profiles. The integrated brightness changes the
molecular excitation throughout the cloud and is visible as 
continuum emission. Moreover, the molecular material in front 
of an \HII region appears partially in absorption.
A large \HII region with electron densities being a factor 10 or 
more higher than in the example above, would thus result in
distinct changes in the lines. Depending on the configuration 
and velocity structure, the CS lines may appear in absorption 
or with P Cygni profiles. Moreover, a strong continuum 
emission would be observed at all frequencies considered.
However, bright \HII regions with high electron densities
are unlikely to be that extended (\cite{Wood}). For 
ultracompact \HII regions the change of the line profiles 
by absorption is negligible due to the small angular size of 
the \HII region so that the molecular excitation is the remaining
effect. From the lack of a bright continuum underlying 
the lines we can, however, exclude configurations with a bright
\HII region here. Weaker compact or ultracompact \HII regions 
-- although possible -- were not
included in the fit computations as they would only influence 
material in their close environment which cannot be resolved 
in this study.}

Regarding the chemical evolution of massive cores one should also
expect a variation of the molecular abundances of CS and C$^{34}$S. 
However, {\changed our present knowledge} is still insufficient to guess reliable
values here (cf. \cite{Bergin}).
As discussed in Sect. \ref{sect_turbulence},
it is also not possible to fit the abundance independently from the
clump filling factor, so that we adopted here a constant CS abundance 
of $4\,10^{-9}$ relative to H$_2$ ignoring any radial variation of the 
abundances. 

\begin{figure}
\epsfig{file=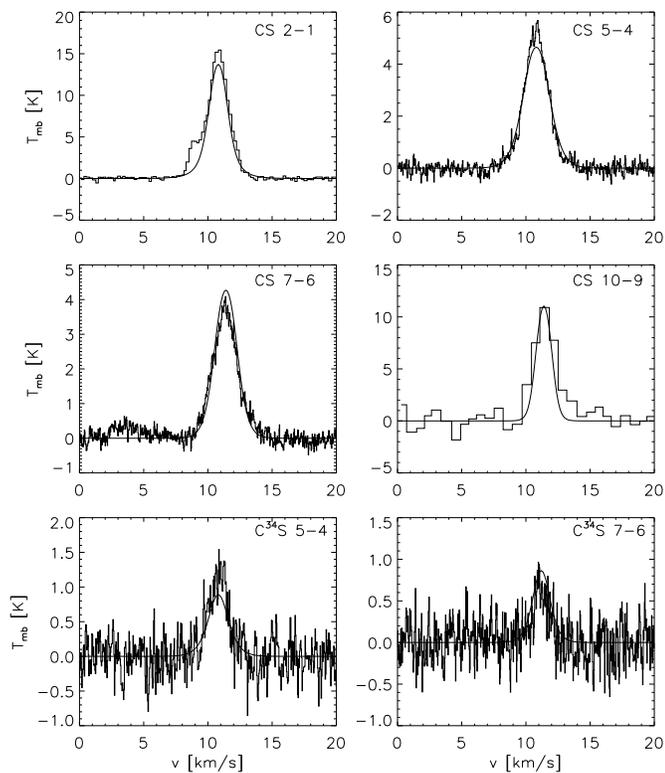, angle=0, width=8.7cm}
\caption{Observed line profiles and best fit model to the 
observations of NGC2024 using a CS to C$^{34}$S abundance
ratio of 13.}
\label{fig_ngc2024}
\end{figure}

An additional test is only possible for NGC2024 where C$^{34}$S is
sufficiently bright so that we could include it
in the fits. Fig. \ref{fig_ngc2024} shows the best fit
model to all six available line profiles.
Equivalent to the results {\changed from the escape probability
model} we get the best match for a relative molecular
abundance $X($CS$)/X($C$^{34}$S$) \approx 13$. All six
lines are simultaneously fitted.

None of the cores except W49A show clearly asymmetric profiles
as a signature of systematic internal velocities.
The weak wings seen {\changed in a few of the other lines} at low
resolution are insufficient to derive any collapse or outflow model.
Thus we have fitted all cores with a static model.

\begin{table*}
\caption{Direct fit parameters and their likely ranges obtained 
in the $\chi^2$ minimisation}
\begin{tabular}{lr@{\,}lr@{\,}lllr@{\,}lrll}
\hline
\multicolumn{1}{c}{Source}
& \multicolumn{2}{c}{ $R\sub{cent}^{\rm max}$}
& \multicolumn{2}{c}{ $R\sub{cloud}^{\rm min}$}
& \multicolumn{1}{c}{ $n\sub{cent}$}
& \multicolumn{1}{c}{ $\alpha_n$}
& \multicolumn{2}{c}{ $T\sub{cent}$}
& \multicolumn{1}{c}{ $\alpha_T$}
& \multicolumn{1}{c}{ $\Delta v\sub{cent}$} 
& \multicolumn{1}{c}{ $\alpha_{\Delta v}$} 
\\
\multicolumn{1}{c}{ }
& \multicolumn{2}{c}{ [pc] } 
& \multicolumn{2}{c}{ [pc] } 
& \multicolumn{1}{c}{ [cm$^{-3}$] } 
&
& \multicolumn{2}{c}{ [K] } 
&
& \multicolumn{1}{c}{ [km/s] } &
\\ \hline
W49A$^{(a)}$     & 0.23 &[:0.50] & 3.6 &[2.2:] & 1.2\,[1.0:1.4]\,10$^7$ & -1.8\,[1.6:2.3] & 92 &[67:125] & 0.0 & 6.6\,[5.1:9.1] & 0.15 \\
W49A$^{(b)}$     & 0.92 &[0.59:1.2] & 3.6\ &[2.6:] & 9.9\,[8.5:11.0]\,10$^5$ & -1.6\,[1.4:1.7] & 109 &[85:138] & 0.0 & 7.1\,[5.6:9.6] & 0.22 \\
W33             & 0.27 &[:0.39] & 1.9 &[0.81:] & 9.1\,[7.0:11.0]\,10$^6$ & -1.9\,[1.1:2.6] & 36 &[30:42] & -0.2 & 4.0\,[2.7:6.2] & 0.44 \\
W51A            & 0.57 &[:0.88] & 3.9 &[2.1:] & 5.3\,[4.3:6.2]\,10$^6$ & -1.6\,[1.2:2.8] & 44 &[35:53] & 0.0 & 8.3\,[6.2:12.0] & 0.23 \\
W3(OH)          & 0.086 &[:0.16] & 2.2 &[0.54:] & 1.2\,[0.99:1.4]\,10$^7$ & -1.7\,[1.4:2.6] & 39 &[32:46] & 0.0& 3.7\,[2.8:5.3] & 0.08 \\
W3              & 0.062 &[:0.10] & 1.4 &[0.30:] & 6.3\,[4.8:7.8]\,10$^6$ & -2.0\,[1.7:2.6] & 64 &[49:79] & -0.2 & 3.0\,[2.1:5.2] & 0.30 \\
S255            & 0.056 &[:0.077] & 1.1 &[0.66:] & 1.6\,[1.4:1.8]\,10$^7$ & -1.5\,[1.3:1.7] & 47 &[41:52] & 0.0 & 2.1\,[1.6:2.9] & 0.14 \\
S235B           & 0.098 &[:0.12] & 0.53 &[0.28:] & 8.3\,[6.8:9.8]\,10$^6$ & -2.1\,[1.5:2.7] & 20 &[17:22] & 0.0 & 1.7\,[1.3:2.3] & 0.56 \\
S106            & 0.024 &[:0.038] & 1.2 &[0.13:] & 8.8\,[7.5:9.9]\,10$^5$ & -1.3\,[1.2:1.4] & 137 &[103:174] & 0.0 & 1.2\,[0.9:1.6] & 0.55 \\
Serpens         & 0.031 &[0.024:0.036] & 0.68 &[0.072:] & 6.4\,[5.6:7.2]\,10$^5$ & -1.6\,[1.5:1.7] & 93 &[74:112] & 0.0 & 2.8\,[2.2:3.6] & 0.47 \\
DR21            & 0.11 &[:0.21] & 1.7 &[0.78:] & 6.5\,[5.3:7.6]\,10$^6$ & -1.5\,[1.3:1.9] & 83 &[61:108] & 0.0 & 1.8\,[1.4:2.5] & 0.52 \\
Mon R2          & 0.097 &[0.057:0.12] & 1.3  &[0.33:] & 1.9\,[1.6:2.1]\,10$^6$ & -1.5\,[1.3:1.8] & 79 &[61:98] & -0.2 & 1.7\,[1.3:2.3] & 0.32 \\
NGC2264         & 0.052 &[:0.085] & 1.2 &[0.31:] & 7.0\,[5.9:8.0]\,10$^6$ & -1.4\,[1.1:2.1] & 37 &[31:44] & 0.0 & 2.9\,[2.2:4.0] & 0.20 \\
OMC-2           & 0.019 &[:0.037] & 0.85 &[0.11:] & 1.4\,[1.2:1.7]\,10$^7$ & -1.5\,[1.1:2.0] & 27 &[24:31] & 0.0 & 1.2\,[0.9:1.5] & 0.38 \\
$\rho$ Oph A    & 0.079 &[0.060:0.10] & 0.68 &[0.16:] & 1.1\,[0.9:1.3]\,10$^6$ & -1.1\,[0.9:1.3] & 20 &[15:24] & 0.0 & 1.2\,[0.9:1.7] & 0.65 \\
NGC2024         & 0.019 &[:0.025] & 0.15 &[0.093:] & 8.0\,[5.8:10.2]\,10$^7$ & -2.2\,[1.9:2.5] &  36 &[31:40] & 0.0 & 1.2\,[0.8:1.8] & 0.48 \\
\hline
\end{tabular}
\label{tab_ltr_results}
\end{table*}

\subsection{Comparing the massive cores}

Tab. \ref{tab_ltr_results} lists the resulting best fit
parameters for all cores with their error intervals. As discussed in 
Sect. \ref{sect_radfunc} the
intervals for the radii are in most cases only limited
at one end. The central density refers to the value 
at the radius $R\sub{cloud}^{\rm min}$.

Comparing the different cores we find that the inner and outer radii
are mainly determined by the selectional bias from the observability as
one massive core. The most distant cores are only detectable
with the KOSMA telescope if they are relatively large
whereas at small distances only small cores are unresolved.

The minimum central density fitting 
the CS lines varies between $10^6$~cm$^{-3}$ for W49A$^{(b)}$ 
and $8\,10^7$~cm$^{-3}$ for NGC2024.
The high value derived for NGC2024 is due to the availability of
the CS 10--9 observations tracing higher densities.
With the three CS lines measured for most cores, only the
density range below about $10^7$ can be reliably traced, so that we
have to take the central density for all clouds
except W49A$^{(b)}$, Serpens, Mon~R2, and $\rho$ Oph~A as lower limits.
The density exponent covers the range between --1.1 and --2.2
where the majority of clouds shows values around --1.6 corresponding 
to large-scale collapse models (Sect. \ref{sect_radfunc}).

Although we don't have a sample where we can expect to set up
statistically significant correlations we can interpret some
general relations. It turns out that the parameters
are not completely independent of each other.  Whereas the 
majority of clouds shows an average temperature between 
20 and 50\,K, the three clouds which needed a 
larger combined abundance factor corresponding to a
higher clump filling factor in the turbulence description (W3, 
S106, Serpens) also tend to require relatively high 
temperatures. There is a clear correlation between the cloud
temperature and the turbulent line width indicating that heating
and turbulent driving might have a related cause.
The cores with a significant temperature exponent also show
a relatively steep density exponent whereas a steep density exponent
itself does not necessarily require a temperature exponent.
These internal relations should be explained from the physical 
nature of the clouds.

There are some peculiarities concerning four massive cores.
The fit of the second component of W49A needs a large inner
region with constant parameters or a very shallow decay of the density 
profile. Here, the main information that we get from the line profiles
is the central density whereas we can hardly constrain the density 
exponent in the power-law range. Regarding the error bars of
the parameters for Serpens we see that the the density structure 
of this relatively nearby cloud is well constrained by its observed 
size whereas the temperature structure is relatively poorly determined. 
In $\rho$ Oph it was not possible 
to fit the line profiles simultaneously with the size of
the smallest core resolved in the observations. This can be explained
by the clumpy structure of the whole region where the excitation
of the core at the central position cannot be treated separately
from the other clumps. Thus we have used for the core model a size 
which is six times larger than the smallest resolved clump and contains 
most of the strong emission observed.
To fit the profiles and spatial extent of Serpens and S106 we have 
to assume a relatively large combined abundance (see Sect. 
\ref{sect_turbulence}) and a low hydrogen density
in the clumps. This peculiarity could be removed when assuming
some foreground CS that increases the apparent size of the massive core
in CS 2--1 but does not contribute to the excitation in the core.

\section{Discussion}

\subsection{Restriction by the observed transitions}

\begin{figure}
\epsfig{file=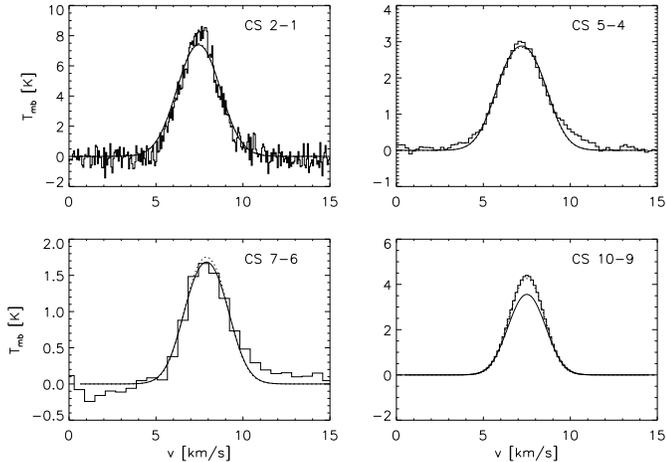, angle=90, width=8.7cm}
\caption{Best fit models to the S255 observations fitting only the three
lower transitions (solid lines) or all four transitions (dotted line).} 
\label{fig_3lines}
\end{figure}

\begin{table}
\caption{Resulting cloud parameters for S255 using either three or
four transitions in the fit.}
\begin{tabular}{lccccccc}
\hline
$n\sub{l}$
& \multicolumn{1}{c}{$\!R\sub{cent}^{\rm max}$}
& \multicolumn{1}{c}{$\!R\sub{cloud}^{\rm min}\!$}
& \multicolumn{1}{c}{$n\sub{cent}$}
& \multicolumn{1}{c}{$\alpha_n$}
& \multicolumn{1}{c}{$\!T\sub{cent}\!$}
& \multicolumn{1}{c}{$\langle n\sub{H_2} \rangle$} 
& \multicolumn{1}{c}{$M\sub{cloud}$} \\
& \multicolumn{1}{c}{[pc]}
& \multicolumn{1}{c}{[pc]}
& \multicolumn{1}{c}{[cm$^{-3}$]} &
& \multicolumn{1}{c}{[K]}
& \multicolumn{1}{c}{[cm$^{-3}$]} 
& \multicolumn{1}{c}{[M$_{\sun}$]} \\
\hline
4 & 0.06 & 1.1 & 1.5\,10$^7$ & -1.5 & 47 & 3600 & 1000 \\
3 & 0.11 & 1.1 & 5.3\,10$^6$ & -1.4 & 49 & 3700 & 1000 \\
\hline
\end{tabular}
\label{tab_3lines}
\end{table}

\begin{table*}
\caption{Error bars for the parameters in S255 using either three or
four transitions in the fit.}
\begin{tabular}{lccccccc}
\hline
$n\sub{l}$ 
& \multicolumn{1}{c}{ $R\sub{cent}^{\rm max}$}
& \multicolumn{1}{c}{ $R\sub{cloud}^{\rm min}$}
& \multicolumn{1}{c}{ $n\sub{cent}$}
& \multicolumn{1}{c}{ $\alpha_n$}
& \multicolumn{1}{c}{ $T\sub{cent}$}
& \multicolumn{1}{c}{ $\Delta v\sub{cent}$} 
& \multicolumn{1}{c}{ $\alpha_{\Delta v}$}\\
& \multicolumn{1}{c}{[pc]}
& \multicolumn{1}{c}{[pc]}
& \multicolumn{1}{c}{[cm$^{-3}$]} &
& \multicolumn{1}{c}{[K]} & 
& \multicolumn{1}{c}{[km/s]} \\
\hline
4 & [:0.08] & [0.66:] & [1.6:2.2]\,10$^7$ & -[1.3:1.7] & [41:52] & [1.6:2.9] & [-0.01:0.42] \\
3 & [:0.21] & [0.68:] & [4.4:6.1]\,10$^6$ & -[1.2:1.8] & [40:59] & [1.7:3.2] & [-0.07:0.70] \\
\hline
\end{tabular}
\label{tab_3errors}
\end{table*}

The main constraint to the core parameters that we can derive is
set by the molecule and the transitions observed. They are only
sensitive to a relatively narrow density range. We can test 
the limitation of the fits provided by the restriction to 
three line profiles by comparing the results obtained for cores 
where we have additional CS 10-9 profiles. With the data
from \cite{Plume} for S255 we investigate how much information
is lost due to the lack of CS 10--9 data in most cores. Fig. 
\ref{fig_3lines} shows the resulting best fits to the central
line profiles in S255 when either all four lines are fitted or only 
the information from the lower three transitions is used. Tab. 
\ref{tab_3lines} lists the corresponding model parameters from the fits. 
Isothermal models provided good fits to the data and
the derived physical parameters are almost identical except 
for the inner radius. The additional information from the 
CS 10--9 transition can set a smaller limit here corresponding
to the higher central densities. When predicting the 10--9 line
data from the best fit of the three other lines
the intensity is too low by only about 20\,\% (Fig. \ref{fig_3lines}). 
Hence, we expect reliable results also for those cores where only 
three lines are observed but it would be favourable to add information
from higher transitions for a better resolution of the densest inner region.

Moreover, a fourth line will reduce the error bars of the parameters. 
In Tab. \ref{tab_3errors} we compare the error obtained 
for the S255 observations using either the three- or the
four-lines fit. As discussed above the possible range of the inner
and outer radii is limited only in one direction. {\changed Thus, the 
fourth line mainly reduces the uncertainty of the inner radius. It hardly
influences the error of the outer radius and the density at the inner radius
but it also reduces the error of the density exponent, the temperature
and the velocity structure.} Thus the inclusion of additional lines
in the model fits would also give a better constraint of the
parameters derived.

\subsection{Comparison of the methods}

\begin{table*}
\caption{Clump parameters derived from the {\it SimLine} results. The
fourth column contains the column density from the escape probability
model (Tab. \ref{tab_epparm}) for comparison and the last column
the distance assumed in the mass derivation}
\begin{tabular}{llrrrrrr}
\hline
\multicolumn{1}{c}{Source}
& \multicolumn{1}{c}{ $\langle n\sub{H_2} \rangle $} 
& \multicolumn{1}{c}{ $\langle N\sub{H_2} \rangle $} 
& \multicolumn{1}{c}{ $N\sub{H_2}$(EP)} 
& \multicolumn{1}{c}{ $\langle T \rangle$}
& \multicolumn{1}{c}{ $M\sub{cloud}$ }
& \multicolumn{1}{c}{ $M\sub{vir}$ }
& \multicolumn{1}{c}{ $D$}
\\
\multicolumn{1}{c}{ }
& \multicolumn{1}{c}{ [cm$^{-3}$] } 
& \multicolumn{2}{c}{ [$10^{23}$~cm$^{-2}$] } 
& \multicolumn{1}{c}{ [K] } 
& \multicolumn{1}{c}{ [M$_{\sun}$] } 
& \multicolumn{1}{c}{ [M$_{\sun}$] } 
& \multicolumn{1}{c}{ [pc] }
\\ \hline
W49A$^{(a)}$     & 1.4\,10$^3$ & 2.6 & 4.6 & 92 & 15000 & 22000 & 11000 \\
W49A$^{(b)}$     & 1.7\,10$^3$ & 0.80 & 4.8 & 109 & 19000 & 14000 & 11000 \\
W33             & 4.6\,10$^3$ & 2.3 & 3.8 & 28 & 7400 & 6300  & 4000 \\
W51A            & 3.9\,10$^3$ & 3.0 & 9.8 & 43 & 54000 & 30000 & 7500 \\
W3(OH)          & 7.7\,10$^2$ & 1.0 & 1.5 & 39 & 1900 & 1900 & 2200 \\
W3              & 2.7\,10$^3$ & 3.6 & 9.7 & 41 & 1700 & 1100 & 2200 \\
S255            & 2.8\,10$^3$ & 1.1 & 0.78 & 47 & 870 & 530 & 2500\\
S235B           & 4.5\,10$^3$ & 0.66 & 0.22 & 20 & 180 & 210  & 1800 \\
S106            & 7.2\,10$^2$ & 0.32 & 0.23 & 137 & 280 & 200 & 600 \\
Serpens         & 7.3\,10$^2$ & 0.22 & 0.17 & 93 & 53 & 80  & 310 \\
DR21            & 1.6\,10$^3$ & 0.82 & 0.79 & 83 & 1800 & 1800 & 3000 \\
Mon R2          & 6.1\,10$^2$ & 0.22 & 0.18 & 54 & 310 & 260 & 950 \\
NGC2264         & 1.1\,10$^3$ & 0.46 & 0.71 & 37 & 450 & 720 & 800 \\
OMC-2           & 6.6\,10$^2$ & 0.34 & 0.24 & 27 & 95 & 52 & 400 \\
$\rho$ Oph A    & 1.3\,10$^3$ & 0.12 & 0.12 & 20 & 96 & 34 & 160 \\
NGC2024         & 2.2\,10$^4$ & 1.3 & 1.6 & 36 & 17 & 35 & 420 \\
\hline
\end{tabular}
\label{tab_coreparams}
\end{table*}

Tab. \ref{tab_coreparams} shows quantities 
characterising the global properties of the clouds
computed from the fit parameters in Tab. \ref{tab_ltr_results}.
The average density in the second column is 
given by the cloud mass within the outer radius.
{\changed We see the strong discrepancy between the average density (Tab.
\ref{tab_coreparams}) and the central clump density 
(Tab. \ref{tab_ltr_results}) reflecting a very inhomogeneous structure
with low volume filling factor of dense clumps.}

In the column 3 and 4 of Tab. \ref{tab_coreparams} one can compare the 
average column density towards the centre in the nonlocal model with 
the column density computed from the escape probability model. {\changed
Here, we have used the molecular column densities from Tab. \ref{tab_epparm}
assuming a CS abundance of $4\,10^{-9}$.} We find an agreement within a factor 
of about 1.5, despite the completely different
analysis applied, except for W49A, W51, and W3 where the column density
from the escape probability model is more than a factor of two higher
and S235B where it is lower. S235B, W49A, and W3 are
the smallest sources in our sample unresolved even in the
CS 2--1 beam. Here, the beam filling factors used in the escape probability
model are uncertain so that they may be responsible for the difference. 
For W51A, no simple explanation 
for the difference is obvious. It is however, by far the most
massive core in our sample so that it might be somewhat peculiar
from that point of view.  In general, we find that the escape probability
analysis provides a reasonable determination for the column density
when we have a good estimate of the beam filling factor. It fails to 
derive correct densities or sizes.

\subsection{Core masses}

Columns 6 and 7 in Tab. \ref{tab_coreparams} shows the core masses
computed in two different ways.
Column 6 gives the integrated mass of the model cloud assuming the 
smallest fitting outer radius and the maximum possible central radius. 
The influence of the inner radius on the total mass is negligible, but the uncertainty
from the lack of information on the outer radius has to be kept in mind.
Increasing the amount of virtually invisible material around the core
by increasing the outer cloud radius can easily increase the total mass
by more than a factor 10. As the mass computation relies on the knowledge of the
CS molecular abundance (Sect. \ref{sect_turbulence}), the resulting
values are to be changed if the true abundances deviate from the
assumed value of $4\,10^{-9}$.

Column 7 contains the core virial mass assuming equipartition
of kinetic and gravitational energy in a homogeneous spherical cloud. We used
the central CS 2--1 line profile and the size of the cloud visible in this 
transition to estimate the velocity dispersion in the line of sight 
and the radius. Following \cite{Lang} 
we obtain the virial mass by
\begin{equation}
M\sub{vir}= 0.0183 \Delta v^2 D \sqrt{\Delta \alpha \Delta\delta}
\end{equation}
where $\Delta v$ is the FWHM of the line (Tab. \ref{tab_lineparams}), 
$D$ the distance of the source, and the $\Delta\alpha$ 
and $\Delta\delta$ the FWHM of the source in declination and 
rectascension in arcmin from Tab. \ref{tab_coresize}.

For all clouds except $\rho$~Oph~A  the agreement between the two masses 
falls within a factor of two. The behaviour of $\rho$~Oph~A is due to the
radius of the excitation model which is larger than the smallest 
resolved core, as discussed above. Hence, it provides a larger mass than
the virial estimate which uses this visible core size. The general
good agreement is quite amazing regarding the uncertainty of the outer boundary 
of the models.  The clouds seem to be virialised and the CS abundance 
estimate holds approximately for all clouds. 

The agreement of the mass from the {\it SimLine} fits with the virial mass
and independent estimates from the literature indicates that the cores are well confined
and our minimum outer radius corresponds to a real, relatively sharp boundary 
for most cores in agreement with the results from continuum 
observations of several cores by Andr\'e et al. (1999). Future
investigations are, however, necessary to confirm this result
because the nature of virialisation is still not understood
and it is therefore not clear how much of the ``invisible'' low
density mass would contribute to the virial mass.

\subsection{Comparison with other observations}

To judge the reliability of the parameters derived here, we can compare
them with core parameters obtained independently from observations in other 
tracers and with other telescopes. In
general they provide only values for few of the cloud parameters
but they may serve as an independent test of our results. We cannot include
a complete discussion of the literature concerning the 15 massive cores 
considered here. Rather we restrict ourselves to a few selected 
observations showing the general power and limitations of the method.

\begin{table}
\caption{Comparison of the NGC2024 FIR5 parameters with results from
{\changed investigations based on other high resolution observations.}}
\begin{tabular}{l@{\hspace*{2mm}}r@{\hspace*{2mm}}r@{\hspace*{2mm}}r@{\hspace*{2mm}}r@{\hspace*{2mm}}r@{\hspace*{2mm}}r}
\hline
 & 
\multicolumn{1}{c}{$R$} &
\multicolumn{1}{c}{$n\sub{cent}$} &
\multicolumn{1}{c}{$N\sub{H_2,cent}$} & 
\multicolumn{1}{c}{$T$}& 
\multicolumn{1}{c}{$M\sub{cloud} $}
\\ & 
\multicolumn{1}{c}{[pc]} &
\multicolumn{1}{c}{$\raisebox{1.5mm}{\Big[}\stackrel{\displaystyle 10^8}{{\rm cm}^{-3}} \raisebox{1.5mm}{\Big]}$} &
\multicolumn{1}{c}{$\raisebox{1.5mm}{\Big[}\stackrel{\displaystyle 10^{24}}{{\rm cm}^{-2}} \raisebox{1.5mm}{\Big]}$} & 
\multicolumn{1}{c}{[K]} & 
\multicolumn{1}{c}{[M$_{\sun}$]} \\
\hline
Mezger$^{\mathrm a}$ & 0.007 & 4.8 & 22 & 19 & 20 \\
Wiesemeyer$^{\mathrm b}$ & $\approx$~0.01 & 0.9-2.3 & 3.1-8.4 & 16-19 & 16-50\\
this paper$^{\mathrm c}$ & 0.15 & $>0.8$ & 9.9 & 36 & 17 \\
\hline
\end{tabular}
\\
${}^{\mathrm a}$ IRAM~30m 870 $\mu$m, SEST 1.3~mm continuum, \cite{Mezger}\\
${}^{\mathrm b}$ IRAM~PdB 3~mm, IRAM~30m 870~$\mu$m, VLA 1.3~cm continuum, \cite{Wiesemeyer}\\
${}^{\mathrm c}$ KOSMA CS and C$^{34}$S, IRAM~30m CS as discussed in the text\\
\label{tab_ngc2024}
\end{table}

For NGC2024 we have compared our data with results from complementary 
high-resolution observations in Tab. \ref{tab_ngc2024}.  \cite{Mezger}
combined the results of SEST observations at 1.3~mm with  IRAM~30m
continuum maps at 870~$\mu$m to identify several clumps in NGC2024
and to deduce their physical properties from the continuum fluxes.
The given values correspond to FIR5 falling at our central position.
\cite{Wiesemeyer}
used a spherically symmetric continuum transfer model to derive the 
physical parameters of FIR5 from 3~mm continuum observations 
taken with the IRAM Plateau de Bure interferometer combined with the 
data from \cite{Mezger} and VLA 1.3~cm observations of \cite{Gaume}.
Depending on the assumed luminosity and dust properties 
they found a range of parameters fitting the observed continuum.
The last line in Tab. \ref{tab_ngc2024} represents the results from 
{\changed the {\it SimLine} fit to our data.}
In contrast to Tab. \ref{tab_coreparams}, the central column density
given here is not averaged but computed towards a central clump to
allow a better comparison with the dust observations which are
able to resolve this clump.

We see that one cannot reveal
the true radius of the core from our low resolution observations.
Moreover, the CS observations cannot trace the same high densities
as the dust observations so that they provide only a lower limit.
It is, however,  already close to the central density given by 
\cite{Wiesemeyer}. The mass and column density derived from our
radiative transfer computations agree quite well with the values
provided by the high-resolution observations. 
A possible explanation for the difference between the gas kinetic 
temperature and the dust temperatures was provided
already by \cite{Schulz}. They performed NH$_3$ and CS observations
of NGC2024 and obtained temperatures between 35 and 40~K at the
position considered. Using a two-component dust model they demonstrated
that these temperatures are also consistent with the observations 
by \cite{Mezger}. Altogether we are able to derive
realistic values for the core parameters with a clumpy radiative
transfer model even if we are not able to deduce the exact object size
as we cannot resolve it. 

W51 consists of three compact molecular cores located within about 
one arcmin.  Interferometric observations by Young et al. (1998) 
and the combination of line and 
continuum measurements by Rudolph et al. (1990) seem to indicate
collapse of the component W51e2 with a mass of about 40000 M$_{\sun}$.
The mass determined by the {\it SimLine} fit is 54000 M$_{\sun}$.
The FCRAO and KOSMA observations show no signatures of collapse as
they are probably blurred by our low spatial resolution.
Sievers et al. (1991) obtained temperatures between
20\,K and 57\,K and Zang \& Ho (1997) derived 40-50~K for an inner 
region of about 0.2 pc and 25-30\,K for the outer cloud based on 
NH$_3$ observations. We were able to fit the observations with an 
isothermal cloud at 44~K, however cannot exclude such a temperature 
structure.  Young et al. (1998) used an LTE code assuming spherical
or spheroidal symmetry to simulate the inner 0.2~pc region of W51e2
fitting the observed ammonia data. They obtained density gradients
of -1.8 to -2.2 somewhat steeper than in our fit, indicating that
a dense central region might be surrounded by an envelope 
with a flatter density gradient. Their central densities between 
1.5 and 22\,10$^6$~cm$^{-3}$ bracket our value of 5\,10$^6$. 
The assumption of a smooth medium  by Young et al. (1998) results
in a central temperature estimate below 25\,K and a steep 
temperature gradient. i.e. significantly lower temperatures than
in our clumpy turbulent model. We have tested this behaviour by trying to
fit the data without clumping in our model and also got low
temperatures below 20~K but quite bad $\chi^2$ values. Thus the
correct treatment of the internal clumping is essential for a 
reliable temperature derivation.

W3(OH) was studied e.g. by Wilson et al. (1991) using VLA observations
of methanol and OH and by Tieftrunk et al. (1998) and \cite{Helmich}
with single dish observations of ammonia and HDO, respectively.
The KOSMA beam covers several maser spots 
and a luminous mm continuum source -- probably a class 0 object. Wilson
et al. (1991) obtained a kinetic temperature of the molecular cores of
20\,K, whereas \cite{Tieftrunk98} derived 27\,K. Both agreed with our
total mass estimate of about 2000 M$_{\sun}$. The radius of 1.3~pc 
determined by Tieftrunk is somewhat smaller than the value computed from
our model, but clearly within the error bar. From the HDO excitation
\cite{Helmich} conclude dust temperatures above 100\,K at densities 
between $10^5$ and $10^6$~cm$^{-3}$, with few embedded clumps at 
$10^7$~cm$^{-3}$. This clump density and the column density 
of about $2\,10^{23}$~cm$^{-2}$ agree approximately with the values from 
our spherical model. The temperature of 
39~K determined from the CS observations falls into the range discussed 
but the difference to the methanol and ammonia values asks for an
explanation. Around the massive
core there is probably still an extended envelope of gas at low densities
insufficient to excite the observed CS transitions (Tieftrunk et al. 1998).

W3 was studied in detail by Tieftrunk et al. (1995, 1997, 1998) using C$^{34}$S,
C$^{18}$O, NH$_3$, and continuum observations. Our beam covers the
two bright components W3 Main and W3 West. Moreover, the region
contains some ultracompact \HII regions related to infrared sources.
The molecular line emission peaks at a position close to W3 West. The line
velocity of our CS observations at -42 km/s agrees with the
velocity of W3 West indicating this component as the main originator
of the observed CS emission. 
The combination of single dish and VLA observations by Tieftrunk et al. 
(1997, 1998) showed extended gas  at a temperature of 25-45\,K, a 
density of 10$^4$~cm$^{-3}$, an size of about 1~pc, and a total 
mass of 1100-1400 M$_{\sun}$. This corresponds well to the parameters
derived from the KOSMA observations. We have
traced the emission to the somewhat larger radius of 1.4~pc, but within
a radius of 1~pc we get about the same average density of
$0.7\,10^4$~cm$^{-3}$. Our mass estimate of 1100~M$_{\sun}$ and the average
temperature of 41~K also agree. {\changed The VLA observations
showed several very compact clumps with a size of 0.02~pc, densities 
of 10$^7$~cm$^{-3}$, and $T\sub{kin}=250$~K. They are not resolvable
from our data, but correspond to the clumps in the turbulence description
and the core parameters derived from
the radiative transfer model show a similar size and density.} In the
smooth temperature distribution assumed in the radiative transfer model
we are not able to resolve hot spots with 250~K but found the
need for an increased temperature towards the centre.

From the group of low mass cores, Castets \& Langer (1995) analysed CS
observations of OMC-2 by means of an LVG analysis providing $T\sub{kin}=24$ K,
a density of $9\,10^5$~cm$^{-3}$ and a CS column density of
$5.4\,10^{13}$ cm$^{-2}$ in agreement with our results. 
{\changed They found already indications for substructure with clump radii
of about 
0.022 pc and higher densities in observations with higher resolution.}
Our analysis shows densities of at least 1.4\,10$^7$~cm$^{-3}$
at a scale of 0.019 pc. The virial mass of 71 $M_{\sun}$ computed
by Castets \& Langer is only somewhat smaller than our mass estimate
of 95 M$_{\sun}$. Recent 1.3~mm observations by \cite{Chini} show at 
least 11 embedded condensations in OMC-2 with masses between 5 and 
8 M$_{\sun}$ and temperatures between 20 and 33 K whereas 350~$\mu{}m$ continuum
data by \cite{Lis98} reveal even 30 clumps but lower temperatures of 
17~K supporting our approach of the clumpy cloud model. 

The comparison shows  that different tracers see different parts 
of a cloud corresponding to different physical conditions. Results
from other authors based on CS observations agree in most cases
quite well, whereas the results from other tracers may considerably
differ. The relatively large uncertainty
in the temperature structure that we cannot resolve within our
analysis asks for additional observations in higher transitions
or at better spatial resolution. For nearby clouds like NGC2024,
OMC-2 or $\rho$ Oph A we get a good agreement with results from 
high-resolution or even interferometric observations, whereas 
for distant massive cores like W49A and W51A there are several 
open question, especially regarding the temperature structure. 
\label{sect_comparison}

\subsection{The physical nature of massive cores}

All massive cores seem to be approximately virialised independent
of their internal structure with respect to the number, distribution
and luminosity of young stars. Although one could
expect that violent bipolar outflows observed in some cores
will drastically change the energy balance in the core, the
physics of the turbulence in the cores seems to be extremely
stable guaranteeing a continuous state of virialisation.

The relatively sharp outer boundary suggested by the mass
estimates can be interpreted in terms of collapse models. The
collapse  of an isothermal sphere would result in a self-similar
density distribution without clear boundary whereas our
results rather tend towards the scenario of a finite-size
Bonnor-Ebert condensation (\cite{Bonnor}). The outer boundary
is, however, not well determined but only set by the mass 
constraints because the radiative transfer model itself cannot 
exclude a continuation of the density structure to larger radii.

The density exponent of about --1.6  derived for most cores
is consistent with several collapse models (see Sect. \ref{sect_radfunc})
but deviations from the exponent for particular clouds up
to values around --2 have to be explained.

On the other hand we have seen that simple collapse models
are not relevant for the massive cores considered here. Clumpiness
is a main feature of all clouds and smooth microturbulent models
are not able to explain the observed lines. In agreement with
other high-resolution observations we find typical clump sizes
of 0.01--0.02~pc at least for the nearby cores. In massive distant
cores the situation might be more complex including a
hierarchy of clump sizes resulting in a larger uncertainty
of the temperature profile derived from our model.

\section{Conclusions}

We have shown that the careful analysis of multi-line single dish 
observations with a relatively large beam can provide a set of
information comparable to single-line interferometric observations.
From a careful excitation analysis using a self-consistent
radiative transfer computation it is possible to deduce
some sub-resolution information. We can infer clump sizes, 
masses and densities at scales below a tenth of the beam size.
However, interferometric observations are necessary to 
determine the exact core geometry including the location and
number of clumps within a dense core. 

The spherically symmetric radiative transfer code used here 
is able to take into account radial gradients in all quantities
and internal clumpiness of the cloud. 
It enables a reliable deduction of the physical parameters
from line profiles observed in sources with a size
close to the spatial resolution limit. 
The method allows to analyse similar observations of objects like
star-forming cores in distant galaxies unresolvable by all today's means.
For a better resolution of the internal temperature structure the 
approach should be combined with sophisticated models on the energy 
balance including the continuum radiative transfer in the future.
Although the simple escape probability analysis gives
a reasonable estimate for the column density, it
fails regarding the density and temperature structure. 

The line analysis shows two essential points:\\ 
{\it i)}
The main constraints on the structural quantities which can be deduced
from the observations are set by the tracer.
The range of densities and temperatures that one can determine
from the radiative transfer calculations is restricted by 
the transitions observed. In case of the CS 2--1, 5--4, and 7--6 lines,
the covered densities range from about $2\,10^5$ to
$10^7$ cm$^{-3}$. With additional information from the
CS 10--9 the upper limit can be extended by another factor 5.
The information from the rarer C${}^{34}$S isotope cannot extend
the density interval but reduces the error bars and provides
better estimates for the clumpiness of the medium.
The high resolution observations discussed in Sects.\ 
\ref{sect_epresults} and \ref{sect_comparison}
show that different tracers provide access to different types 
of information whereas the parameters from our CS observations
agree well with the CS results there. \\
{\it ii)} Temperature and clumpiness are related quantities. When 
turbulent clumping in the cloud is neglected, the temperature
determination will necessarily fail. On the other hand does 
accurate information on the clumpy structure of a massive core help to
constrain the temperature structure. Additional observations in 
higher transitions or complementary estimates of the clumpiness 
will help to reduce the uncertainty of the temperatures. Thus
spatial resolution is still essential. For nearby clouds 
we get a good agreement with results from other high-resolution
observations, but for distant massive cores the temperature structure
is still an open question.

All massive cores that we have analysed are characterised by turbulent
clumpiness with typical clump sizes of 0.01--0.02~pc.
The clouds are approximately virialised and show density gradients
around --1.6 but with a scatter between --1.1 and --2.2.
Large parts of the cores follow a constant temperature but we must
admit a considerable uncertainty in the most inner and outer parts. 
The correlation between the cloud temperature and the turbulent line 
width indicates that related processes should be responsible for
heating and turbulent driving.

Future observations of dense cores should focus on different tracers
to gain access to additional information which cannot be deduced
from a single tracer such as CS. As a drawback, 
the full uncertainty of today's chemical models 
will enter and partially limit the interpretation of the observations.

\begin{acknowledgements}
We thank J. Howe for providing us with the CS 2--1 observational data.
We are grateful to the anonymous referee for many detailed
comments helping to improve the paper considerably.
This project was supported by the \emph{Deut\-sche
For\-schungs\-ge\-mein\-schaft} through the grant SFB 301C.
The KOSMA 3\,m radio telescope at
Gornergrat-S\"ud Observatory is operated by the University of
Cologne and supported by the Land Nordrhein-Westfalen. The receiver
development was partly funded by the Bundesminister f\"ur Forschung
and Technologie. The Observatory is administered by the
Inter\-natio\-nale Stif\-tung Hoch\-alpine Forschungs\-stationen
Jung\-frau\-joch und Gornergrat, Bern.
The research has made use of NASA's Astrophysics Data System
Abstract Service.
\end{acknowledgements}

\appendix
\section{The radiative transfer problem in 
the escape probability approximation}

\subsection{The general radiative transfer problem}

The physical parameters of a cloud model and the
emerging line profiles and intensities are linked by the
radiative transfer problem. It relates the molecular emission
and absorption coefficient at one point to the
radiation field determined by the emission and transfer of 
radiation at other locations in the cloud.
The quantity entering the balance equations for the level populations at
a point $\vec{r}$ is the local radiative energy density $u$ within
the frequency range for each transition:
\begin{eqnarray}
u(\vec{r})&=&\int_{4\pi} \! d\Omega\; u(\vec{r},\vec{n}) \nonumber \\
\quad u(\vec{r},\vec{n})&=& {1 \over c} \int_{-\infty}^{\infty} d\nu\; 
I_\nu(\vec{r},\vec{n}) \Phi_\nu (\vec{r},\vec{n})\;.
\label{ufirst}
\end{eqnarray}
Here, $u(\vec{r},\vec{n})$ is the absorbable radiative energy coming from direction
$\vec{n}$ and $I_\nu(\vec{r},\vec{n})$ is the intensity at a given frequency
within this direction. It is determined by the radiative transfer 
equation
\begin{equation}
\vec{n} \nabla I_\nu(\vec{r},\vec{n}) = - \kappa_\nu(\vec{r},\vec{n})
I_\nu(\vec{r},\vec{n}) + \epsilon_\nu(\vec{r},\vec{n})
\label{eq_transfer}
\end{equation}
Assuming complete redistribution the profile for the absorption and the
emission coefficients, $\kappa_\nu(\vec{n},s)$ and $\epsilon_\nu(\vec{n},s)$
is given by the same local line profile $\Phi_\nu$. For Maxwellian velocity
distributions it is a Gaussian:
\begin{equation}
\Phi_\nu(\vec{n},\vec{r})={1 \over \sqrt{\pi} \sigma} 
\exp\left(-{(\nu-\vec{n}\vec{v}(\vec{r}))^2\over \sigma^2}\right)\;.
\end{equation}
Here, $\vec{v}$ is the velocity of the local volume element
written in units of the frequency $\nu$. The frequently used
FWHM of the distribution is related to $\sigma$ by FWHM$=2\sqrt{\rm ln\,2}\,\sigma$.

For a molecular cloud this results in a huge system of integral
equations interconnecting the level populations and intensities
at all points within a cloud.

\subsection{The escape probability model}

A simple way to avoid the nonlinear equation system is the escape 
probability approximation that is widely applied to interpret 
molecular line data.
It is based on the assumption that the excitation, and thus the 
absorption and emission coefficients, are constant within
those parts of a cloud which are radiatively coupled.
Then the radiative transfer equation (Eq. \ref{eq_transfer}) can be 
integrated analytically. We obtain for the integrated radiative 
energy density:
\begin{equation}
u(\vec{n})={1 \over c} \left [\beta(\vec{n}) I\sub{bg}(\vec{n})+\left(1-\beta(\vec{n})
\right) {\epsilon_l \over \kappa_l}\right]\;.
\label{eq_udir}
\end{equation}
where the subscript $l$ denotes line integrated quantities.
The term $\beta(\vec{n})$ is the probability that a photon 
can escape or penetrate along the line of sight $\vec{n}$ from the
considered point to the boundary of the interaction region or vice
versa.
\begin{eqnarray}
\beta(\vec{n}) &= & \int_{-\infty}^{\infty} d\nu\; \Phi_\nu(\vec{r},\vec{n})
\times \exp (-\tau_\nu(\vec{r},\vec{n}))\\
{\rm with} && \tau_\nu(\vec{r},\vec{n})=\int_{-\infty}^{\vec{r}} ds_{\vec{n}}\;
\kappa_l(\vec{r}) \Phi_\nu(\vec{r},\vec{n})
\end{eqnarray}
where the integration path $ds_{\vec{n}}$ follows the direction $\vec{n}$.

There are two main concepts to define the interaction region and thus to
compute the escape probability. The first one is the large velocity
gradient approximation introduced by \cite{Sobolev}. Here,
the interaction region is determined by a velocity gradient in the cloud
that displaces the line profiles along the line of sight so that
distant regions are radiatively decoupled. When the resulting interaction
region is sufficiently small it is justified to assume constant parameters.
The escape probability then follows
\begin{eqnarray}
\beta(\vec{n}) 
&=& \left. \left[1-\exp \left( -\tau\sub{LVG}
\right)\right]\right/ \tau\sub{LVG}\\
{\rm with} && \tau\sub{LVG}={\kappa_l \over |(\vec{n}\nabla) (\vec{n}\vec{v})|}
\label{eq_lvg}
\end{eqnarray}
(c.f. \cite{Ossenkopf97}).
The radiative energy density in each direction is determined
by two local quantities only: the source function $S=\epsilon_l/\kappa_l$
and the optical depth of the interaction region $\tau\sub{LVG}$.
The observable brightness temperature at the line centre is
constant over all regions with the same velocity gradient
\begin{equation}
T\sub{B} \approx {c^2 \over 2 k \nu^2} \left( S - I\sub{bg}\right)
\left[1-\exp(-2\tau\sub{LVG})\right]
\label{eq_tmb_lvg}
\end{equation}

In  molecular clouds, the local velocity gradients
are unknown, however. It is generally assumed that
the total observed line width, which is composed from turbulent, thermal, and 
systematic contributions, can be used as the measure of the velocity gradient
over the cloud size. This approach was applied by \cite{Plume} for
the massive cores discussed in the text.

We used another method, the static escape probability
model. It does not depend on the velocity structure but 
assumes a special geometry of 
the interaction region. \cite{Stutzki85} solved the problem for
a homogeneous spherical cloud with constant excitation parameters.
The resulting escape probability is taken to be constant
\begin{equation}
\beta=\int_{-\infty}^{\infty} d\nu\; \Phi_\nu \times \exp (-\tau_\nu)\;.
\label{eq_beta_ep}
\end{equation}
where $\tau_\nu$ is the optical depth at the cloud centre.

The surface brightness temperature towards the centre of the cloud is
given by the same expression as Eq. (\ref{eq_tmb_lvg}) when we use
the line integrated optical depth at the cloud centre $\tau_l$ instead of
$\tau\sub{LVG}$. It decays with growing distance from the cloud
centre. Averaged over the whole cloud, the brightness temperature 
at the line centre is given by
\begin{eqnarray}
T\sub{B} & \approx&  {c^2 \over 2 k \nu^2} \left( S - I\sub{bg}\right)
\left(1-e(-2\tau_l)\right) \label{eq_tmb_ep} \label{eq_tmb_esc}
\label{eq_tmb_statep}\\
{\rm with} & & e(x)={2 \over x^2} \left( 1 - \exp(-x)(1+x)\right) \nonumber \;.
\end{eqnarray}
This value would be observed with a beam much larger than the cloud. 

When the velocity gradient in the LVG approximation is computed from
the total line width and the cloud size, it turns out that both methods
agree when applied to observations with a small beam towards the cloud
centre. Only for large-beam observations, they differ in the functions 
in Eq. (\ref{eq_tmb_lvg}) and (\ref{eq_tmb_ep}), which are either 
$\exp(-2\tau)$ or $e(-2\tau)$, but result in similar values.

By setting up a table of beam temperatures from Eqs. (\ref{eq_tmb_lvg})
and (\ref{eq_tmb_ep}) and comparing the observed line intensities
with the tabulated values we can derive three parameters from the
observations: the kinetic 
temperature $T\sub{kin}$ and the gas density $n\sub{H_2}$ providing 
mainly the source function, and the column density of the considered 
molecules relative to the line width $N/\Delta v$ providing the photon 
escape probability. 

\section{SimLine - A one-dimensional radiative transfer code}

\subsection{The radiative transfer problem}

{\it SimLine} solves the line radiative transfer problem discussed
in Appx.\ A.1 in a spherically symmetric configuration by means
of a $\lambda$-iteration. The code is similar to the concept described by
\cite{Dickel} but it contains several extensions and achieves
a higher accuracy from an adaptive discretisation of all independent
quantities. 

{\it SimLine} integrates the radiative transfer equation (\ref{eq_transfer}) 
for a number of rays numerically.
In spherical symmetry it is sufficient to consider the propagation
of radiation in one arbitrary direction which is taken as $z$ here. 
The integral is computed stepwise from $z_{i-1}$ to $z_{i}$
\begin{eqnarray}
I_\nu(p,z_i) &=& \exp \left( -\int_{z_{i-1}}^{z_i} \!\!\! dz\;
\kappa_\nu(p,z) \right) \Bigg[ I_z(\nu,p,z_{i-1}) \nonumber \\ & 
+&\left. \int_{z_{i-1}}^{z_i} \!\!\! dz\;\epsilon_\nu(p,z) \exp \left( \int_{z_{i-1}}^z
\!\!\! dz' \kappa_\nu(p,z') \right) \right]
\label{transfer}
\end{eqnarray}
where $p$ denotes the displacement variable perpendicular to the
$z$ direction. To minimise the discretisation error the
integral does not use the source function but only 
the emission and absorption coefficients which are linear in
the level populations. They are assumed to change linearly between
the grid points and the exact integration formula for a linear behaviour 
(which is not given here, but can be derived straight forward) is applied.
This approach provides a reduction of the integration error to third order.
The radial grid is dynamically adjusted to give a maximum variation
of the level populations between two neighbouring points below
a certain limit. In case of strong velocity gradients additional
points are included on the $z$-scale for a sufficiently dense sampling
of changes in the profile function.
The incident radiation at the outer boundary of the cloud is
assumed to follow a black body spectrum.

In spherical symmetry the spatial integration of the radiative energy
density (Eq. \ref{ufirst}) can be reduced to
\begin{eqnarray}
u(r) &=& {2\pi\over r c} \int_{-r}^r dz' \int_{-\infty}^\infty d\nu\;
\Phi_\nu(p',z') I_\nu(p',z') \\
{\rm with}&& p'=\sqrt{r^2-z'^2} \nonumber
\label{uzint} 
\end{eqnarray}
The grid of rays tangential to the radial grid 
is refined by additional rays at intermediate $p'$ values
to guarantee a sufficiently dense sampling on the $z'$ scale
The integration uses a cubic spline interpolation.

With the values of the radiative energy density at each radial point and for
each transition, the system of balance equations can be solved
providing new level populations. Here, a LU decomposition algorithm
with iterative improvement (\cite{Press}) is used. The new level populations
are used in the next iteration as input for the radiative transfer equation.
The whole $\lambda$-iteration scheme is 
solved using the convergence accelerator introduced by \cite{Auer87}.

Depending on the physical situation the initial guess is either
the optically thin limit, thermalisation or the solution
of the radiative transfer equation using the LVG approximation (Eq.
\ref{eq_udir}). The stability of the local radiation field is used
as convergence criterion. The number of iterations required for 
convergence depends strongly on the optical depth of the model 
cloud. For the examples discussed in this
paper only about a dozen iterations were necessary but other test cases
with complex molecules like water, {\changed non-monotonic} velocity gradients,
and high optical depths require several hundred iterations.

\subsection{The local turbulence approximation}

The turbulence description uses two additional parameters for 
each spatial point: the width of the velocity distribution 
$\sigma$ providing the
local emission profile for optically thin lines and the correlation length
of the macroturbulent density or velocity distribution
$l\sub{corr}$.

The width of the velocity distribution $\sigma$ is composed of a 
turbulent and a thermal contribution
\begin{equation}
\sigma = {\nu_0 \over c} \,\sqrt{ {2kT\sub{kin} \over m\sub{mol}} + {2 \over 3} 
\langle v\sub{turb}^2 \rangle}
\end{equation}
where a Maxwellian distribution of turbulent velocities is assumed. 
The relation between the FWHM and the variance of the turbulent
velocity distribution is given by 
${\rm FWHM}(v\sub{turb})=\sqrt{8/3\times{\rm ln}2\;
\langle v\sub{turb}^2 \rangle}$). 
The long range variation of the turbulence spectrum as described by means of
a Kolmogorov or Larson exponent is simulated by a radially varying 
turbulent velocity dispersion 
$\sqrt{\langle v\sub{turb}^2\rangle} \propto r^\gamma$. Exponents
$\gamma$ between about 0.1 (\cite{Goodman}) and 0.7 (Fuller \& Myers 1992)
are observationally justified.

For the local treatment of coherent units in a turbulent medium the
considered volume element is subdivided into numerous
clumps with a thermal internal velocity dispersion.
When each clump is characterized by a Gaussian density distribution
of molecules with about the same velocity
$ n(r)=n_0\times\exp(-r^2/r\sub{cl}^2) $
the effective absorption coefficient at the considered
velocity for the whole medium is
\begin{equation}
\kappa\sub{eff}=n\sub{cl} \times \pi r\sub{cl}^2 \int_0^{\tau\sub{cl}}
{1-\exp(-\tau) \over \tau} d\tau
\end{equation}
where $n\sub{cl}$ is the number density of contributing cells and 
$\tau\sub{cl}=\sqrt{\pi}\,\kappa r\sub{cl}$ is their central opacity
(Martin et al. 1984). 
As the clumps size $r\sub{cl}$ is the length on which the abundance 
of molecules within the same thermal velocity profile 
is reduced by the factor $1/e$, we can compute it from the 
correlation length of the velocity or density structure by 
$ r\sub{cl}=l\sub{corr}\times \sigma\sub{th} /\sigma$. 

When the turbulent velocity dispersion $\sigma$ is at least three times 
as large as the thermal velocity dispersion $\sigma\sub{th}$, we obtain
an effective absorption coefficient
\begin{equation}
\kappa\sub{eff,\nu}=n\sub{ges} \pi r\sub{cl}^2 \times A(\tau\sub{cl})
\times {\sigma\sub{th} \over \sigma} \exp{}\left(-{(\nu-\nu_0)^2
\over \sigma^2}\right)
\label{eq_keff}
\end{equation}
with 
\begin{equation}
A(\tau)={1\over \sqrt{\pi}}\int_{-\infty}^{\infty}dv
\int_0^{\tau \exp(-v^2)} {1-\exp(-\tau') \over \tau'} d\tau'
\end{equation}
Here, $n\sub{ges}$ is the total number density of clumps.
In case of incompressible turbulence, i.e. clumping in velocity space, 
it is equal to the reciprocal cell volume. For small values of the
clump opacity,
$A(\tau\sub{cl})$ is identical to $\tau\sub{cl}$ and we reproduce
the microturbulent limit. For large $\tau\sub{cl}$, 
the function $A(\tau\sub{cl})$ saturates and we obtain a significant
reduction of the effective absorption coefficient.
In case of density clumps, $\kappa\sub{eff}(\nu)$ is further reduced
by the filling factor entering $n\sub{ges}$. In {\it SimLine}, this 
is simulated by a corresponding artificial reduction of the molecular 
abundance.

\subsection{The central \HII region}

To simulate the effect of a central continuum 
source in the cloud, it is possible to assume an \HII region in the cloud 
centre.
The \HII region is characterised by two parameters, the electron density
$n\sub{e}$ and the kinetic electron temperature $T\sub{e}$.

The absorption coefficient for electron-ion bremsstrahlung in the
Rayleigh-Jeans approximation is given by:
\begin{equation}
\kappa_\nu={8 \over 3\sqrt{2\pi}}\; {e^6 \over (4\pi\epsilon_0 m\sub{e})^3 c}
\left(n\sub{e} \over \nu\right)^2 \left(m\sub{e}\over k T\sub{e}\right)^{3/2}
{\rm ln}\Lambda
\label{eq_hii_kappa}
\end{equation}
where it is assumed that the gas is singly ionised and $\Lambda$ is 
given by
\begin{equation}
\Lambda=\left(2k T\sub{e} \over \gamma m\sub{e}\right)^{3/2} 
{4\pi \epsilon_0 m\sub{e} \over \pi\gamma e^2 \nu} \approx
4.96\,10^7 \left(T\over {\rm K}\right)^{3/2} {{\rm Hz} \over \nu}
\end{equation}
for $T\sub{e}<3.6\,10^5$K. The quantities $e$ and $m\sub{e}$ denote the
electron charge and mass, $c$ is the vacuum light velocity, and
$\gamma=1.781$ (Lang 1980).

For a thermal plasma, the emission coefficient follows
from the Planck function
\begin{equation}
\epsilon_\nu=\kappa_\nu\times B_\nu(T\sub{e})\;.
\label{eq_hii_ep}
\end{equation}

In the radiative transfer computations the frequency dependence 
of these continuum coefficients is neglected within the molecular line 
width. Within the \HII region, we substitute the molecular coefficients
in Eq. (\ref{transfer}) by the quantities from Eqs. (\ref{eq_hii_kappa})
and (\ref{eq_hii_ep}) so that we locally switch to a continuum radiative
transfer.
\label{sect_hii_region}

\subsection{Computation of beam temperatures}

When the level populations are known, the beam temperature relative 
to the background is computed by the convolution of the emergent intensity
with the telescope beam.
\begin{equation}
T\sub{mb}={c^2 \over 2 k \nu_0^2} \;{\displaystyle \int_0^{2\pi} \!\!d\phi
\int_0^\infty \!\!dp \;p (I_\nu^S(p)-I\sub{bg} )
f\sub{mb}(p,\phi) \over \displaystyle
\int_0^{2\pi} \!\!d\phi \int_0^\infty \!\!dp\;
p f\sub{mb}(p,\phi)}
\label{beam}
\end{equation}
The intensity $I_\nu^S(p)$ is the value on the cloud
surface $I_\nu^S(p)=I_\nu(p,\sqrt{R\sub{cloud}^2-p^2})$.
We assume a Gaussian profile for the telescope beam
\begin{equation}
f\sub{mb}(p,\phi)= 
\exp{}\left(-(p-p\sub{offset})^2(1+\phi^2)^2 \over \sigma\sub{mb}^2\right)
\end{equation}
The projected beam width is computed from the angular
width by $\sigma\sub{mb} = \pi/648000\, D\, \sigma\sub{mb}['']
= 2.912\,10^{-6}\, D\, {\rm FWHM} ['']$
where $D$ is the distance of the cloud. The program
computes a radial map with arbitrary spacings.

\subsection{The general code design}

The design of the code is directed towards a high
accuracy of the computed line profiles. All errors in
the different steps of the program are explicitly user 
controlled by setting thresholds. All discretisations
necessary to treat the problem numerically are performed in
an adaptive way, i.e. there is no predefined grid 
and all grid parameters will change during the iteration procedure.
The system of balance equations is truncated whenever the excitation 
of all higher levels falls below a chosen accuracy limit.

Furthermore, the code was pushed towards a high flexibility,
i.e. the ability to treat a very broad range of physical
parameters with the same accuracy and without numerical
limitations. The systematic velocities e.g. may range from
0 to several times the turbulent velocity and the optical depths 
may vary from negative values for weak masing to values of
several thousands.

The program is not optimised towards a high speed. Other codes 
with lower inherentaccuracy may easily run a factor 10 faster
and further improvements are possible. Nevertheless,
the code is suitable for an interactive work even on a small
PC with execution times of a few seconds for the models 
considered in this paper.


\begin{thebibliography}{}

\bibitem[Andr\'e et al. 1999]{Andre}Andr\'e P., Bacmann A., Motte F., 
        Ward-Thompson D. 1999, in Ossenkopf V., Stutzki J., Winnewisser G.
        (eds.) The Physics and Chemistry of the Interstellar Medium, 
        GCA-Verlag Herdecke, p. 241
\bibitem[Auer (1987)]{Auer87}Auer L.H. 1987, 
        In: Kalkofen W. (Ed.), Numerical Radiative Transfer,
        Cambridge Univ. Press, p. 101
\bibitem[Basu \& Mouschovias 1995]{Basu}Basu  S., Mouschovias T.C. 1995,
        ApJ 452,  386  
\bibitem[Basu \& Mouschovias 1995]{Basu95}Basu  S., Mouschovias T.C. 1995,
        ApJ 453,  271  
\bibitem[Batrla et al. (1983)]{Batrla}Batrla W., Wilson T.L., Ruf K., 
        Bastien P. 1983,  A\&A 128,  279  
\bibitem[Bodenheimer \& Sweigart (1968)]{Bodenheimer}Bodenheimer P., 
        Sweigart A. 1968, ApJ 152, 515
\bibitem[Bergin \& Langer 1997]{Bergin}Bergin E.A., Langer W.D. 1997, ApJ 486, 316
\bibitem[Bonnor 1956]{Bonnor}Bonnor W.B. 1956, MNRAS 116, 351
\bibitem[Castets \& Langer (1995)]{Castets} Castets  A., Langer W.D. 1995,
        A\&A 294,  835  
\bibitem[Cesaroni \& Wilson (1994)]{Cesaroni}  Cesaroni R., Wilson T.L. 1994,  
        A\&A 281,  209  
\bibitem[Chini et al. (1997)]{Chini}Chini R. Reipurth B., Ward-Thompson D.,
        Bally J., Nyman L.-A., Sievers A., Billawala Y. 1997, ApJ 474, L135
\bibitem[Dickel \& Auer (1994)]{Dickel} Dickel H.R., Auer L.H. 1994, ApJ 437,222
\bibitem[Fuller \& Myers (1992)]{Fuller}Fuller G.A., Myers P.C. 1992,
        ApJ 384, 523
\bibitem[Galli \& Shu 1993]{Galli}  Galli  D., Shu F.H. 1993,  ApJ
         417,  243 
\bibitem[Galli et al. 1999]{Galli99} Galli D., Lizano S., Li Z.-Y., Adams F.C., 
        Shu F.H. 1999, ApJ 521, 630
\bibitem[Garay \& Lizano 1999]{Garay}
	Garay G., Lizano S. 1999, PASP 111, 1049
\bibitem[Garden \& Carlstrom 1992]{Garden}  Garden R.P., Carlstrom J.E. 
        1992,  ApJ 392, 602
\bibitem[Gaume et al. (1992)]{Gaume} Gaume R.A., Johnston K.J., Wilson T.L. 
        1992, ApJ 388, 489
\bibitem[Giannakopoulou et al. (1997)]{Giannakopoulou}  Giannakopoulou  J., 
        Mitchell G.F., Hasegawa T.I., Matthews H.E., Maillard J. 1997,  
        ApJ 487,  346  
\bibitem[Goldsmith \& Mao (1983)]{Goldsmith}  Goldsmith P.F., Mao X.-J. 1983,  
        ApJ 265, 791  
\bibitem[Goodman et al. 1998]{Goodman}Goodman A.A., Barranco J.A.,
        Wilner D.J., Heyer M.H. 1998, ApJ 504, 223
\bibitem[Haschick \& Ho (1983)]{Haschick} Haschick A.D., Ho P.T.P. 1983,  
        ApJ  267, 638  
\bibitem[Hatchell et al. (1998)]{Hatchell}Hatchell J., Thompson M.A., Millar T.J.,
        Macdonald G.H. 1998, A\&A 338, 713
\bibitem[Helmich et al. (1996)]{Helmich}Helmich F.P., van Dishoeck E.F.,
        Jansen D.J. 1996, A\&A 313, 657
\bibitem[Ho et al. (1993)]{Ho}  Ho P.T.P, Peng Y., Torrelles J.M., Gomez
        J.F., Rodriguez L.F., Canto J. 1993,  ApJ 408,  565  
\bibitem[Hogerheijde et al. (1999)]{Hogerheijde}Hogerheijde M.R., van Dishoeck E.F., 
        Salverda J.M., Blake G.A. 1999, ApJ 513, 350
\bibitem[Jaffe et al. (1984)]{Jaffe}  Jaffe D.T., Davidson J.A., Dragovan M.,
        Hildebrand R.H. 1984,  ApJ 284, 637  
\bibitem[Jackson & Kraemer (1994)]{Jackson}Jackson J.M., Kraemer K.E. 1994, ApJ 429, L37
\bibitem[Keto (1990)]{Keto} Keto E.R. 1990, ApJ 350, 722
\bibitem[Keto & Ho (1989)]{KetoHo} Keto E.R., Ho P.T.P. 1989, ApJ 347, 349
\bibitem[Kr\"ugel et al. (1987)]{Kruegel}  Kr\"ugel  E., G\"usten R., Schulz
        A., Thum C. 1987,  A\&A 185,  283  
\bibitem[Lada et al. (1991)]{Lada91}  Lada E.A., Bally J., Stark A.A. 1991,
        ApJ  368,  432-444  
\bibitem[Lada et al. (1997)]{Lada97}  Lada E.A., Evans N.J., Falgarone E. 1997,
        ApJ 488,  286  
\bibitem[Lang (1980)]{Lang} Lang K.R. 1980, Astrophysical Formulae, 
        Springer-Verlag Berlin, p.553
\bibitem[Larson 1969]{Larson69}Larson R.B. 1969, MNRAS 145, 271
\bibitem[Li \& Shu 1997]{Li}Li Z.-Y., Shu F.H. 1997, ApJ 475, 237
\bibitem[Lis et. al. (1998)]{Lis98}Lis D.C., Keen J., Dowell C.D., 
	Benford D.J., Phillips T.G., Hunter T.R., Wang N. 1998, ApJ 509, 299
\bibitem[Liseau et al. (1995)]{Liseau}  Liseau R., Lorenzetti D., Molinari 
        S., Nisini B., Saraceno P., Spinoglio L. 1995 ,  A\&A 300, 493  
\bibitem[Martin et al. (1984)]{MSH}Martin H.M., Sanders D.B., Hills R.E.
        1984, MNRAS 208, 35
\bibitem[McMullin et al. (1994)]{McMullin}  McMullin J.P., Mundy L.G., 
        Wilking B.A., Hezel T., Blake G.A. 1994,  ApJ 424,  222  
\bibitem[Mezger et al. (1992)]{Mezger}  Mezger P.G., Sievers A.W., Haslam 
        C.G.T., Kreysa E., Lemke R., Mauersberger R., Wilson T.L. 1992,  
        A\&A 256,  631 
\bibitem[Miesch \& Bally (1994)]{Miesch}Miesch M.S., Bally J. 1994, ApJ 
        429, 645
\bibitem[Montalban et al. (1990)]{Montalban}  Montalban  J., Bachiller 
        R., Martin-Pintado J., Tafalla M., Gomez-Gonzalez J. 1990,  
        A\&A 233, 527  
\bibitem[Mundy et al. (1986)]{Mundy}Mundy L.G., Evans N.J., Snell R.L., 
        Goldsmith P.F., Bally J. 1986, ApJ 306, 670
\bibitem[Myers 1999]{Myers} Myers P.C. 1999, 
        in Ossenkopf V., Stutzki J., Winnewisser G.
        (eds.) The Physics and Chemistry of the Interstellar Medium, 
        GCA-Verlag Herdecke, p. 227
\bibitem[Nakano \& Yoshida (1986)]{Nakano} Nakano M., Yoshida S. 1986,
        PASJ 38, 531
\bibitem[Plume et al. (1997)]{Plume}  Plume  R., Affe D. T., Evans N.J.,
        Martin-Pintado J., Gomez-Gonzalez J. 1997,  ApJ 476,  730
\bibitem[Press et al. 1992]{Press} Press W.H., Teukolsky S.A., Vetterling W.T., 
        Flannery B.P. 1992, Numerical Recipies in FORTRAN, Cambridge 
        Univ. Press, p. 402
\bibitem[Ossenkopf 1997]{Ossenkopf97}Ossenkopf V. 1997, New Astron.
        2, 365
\bibitem[Roberts et al. (1995)]{Roberts}  Roberts  D.A., Crutcher R.M., 
        Troland T.H. 1997,  ApJ 442, 208  
\bibitem[Rudolph et al. (1990)]{Rudolph}
	Rudolph A., Welch W.J., Palmer P., Dubrulle B. 1990, ApJ 363, 528
\bibitem[Roberts et al. (1997)]{Roberts97}  Roberts  D.A., Crutcher R.M., 
        Troland T.H. 1997,  ApJ 479, 318  
\bibitem[Schneider et al. (1999)]{Schneider}Schneider N., Simon R., Kramer C.,
        Stutzki J., Winnewisser G. 1999, in Ossenkopf V., Stutzki J., Winnewisser G.
        (eds.) The Physics and Chemistry of the Interstellar Medium, 
        GCA-Verlag Herdecke, p. 128
\bibitem[Schulz et al. (1991)]{Schulz}Schulz A., G\"usten R., Zylka R., 
	Serabyn E. 1991, A\&A 246, 570
\bibitem[Scoville \& Kwan (1976)]{Scoville}Scoville N.Z., Kwan J. 1976, 
        ApJ 206, 718
\bibitem[Shu (1977)]{Shu}Shu F.H. 1977, ApJ 214, 488
\bibitem[Sievers et al. (1991)]{Sievers}  Sievers  A.W., Mezger P.G., 
        Kreysa E., Haslam C.G.T., Lemke R., Gordon M.A. 1991,  A\&A 251,  231  
\bibitem[Sobolev (1957)]{Sobolev}Sobolev V.V. 1957, Soviet. Astron. 1, 678
\bibitem[Stutzki \& Winnewisser (1985)]{Stutzki85}Stutzki J., Winnewisser G.
        1985, A\&A 144, 13
\bibitem[Stutzki et al. (1998)]{stutzki}Stutzki J., Bensch F., 
        Heithausen A., Ossenkopf V., Zielinsky M. 1998, A\&A 336, 697 
\bibitem[Tieftrunk et al. (1995)]{Tieftrunk95} Tieftrunk A.R., Wilson T.L., 
        Steppe H., Gaume R.A., Johnston K.J., Claussen M.J. 1995,  A\&A 
        303,  901 
\bibitem[Tieftrunk et al. (1997)]{Tieftrunk97} Tieftrunk A.R., Gaume R.A.,
        Claussen M.J., Wilson T.L., Johnstone K.J. 1997, A\&A 318, 931
\bibitem[Tieftrunk et al. (1998)]{Tieftrunk98} Tieftrunk A.R., Gaume R.A., 
        Wilson T.L. 1998,  A\&A  340,  232 
\bibitem[Turner et al. (1992)]{Turner}Turner B.E., Chan K.-W., Green S.,
        Lubowich D.A.1992, ApJ 399, 114
\bibitem[Welch et al. (1987)]{Welch} Welch W.J., Dreher J.W., Kackson J.M.,
        Terebey S., Vogel S.N. 1987, Science 238, 1550
\bibitem[White 1977]{White} White R.E. 1977, ApJ 211, 744
\bibitem[Wiesemeyer et al. (1997)]{Wiesemeyer}Wiesemeyer H., G\"usten R.,
        Wink J.E., Yorke H.W 1997, A\&A 320, 287
\bibitem[Wilson et al. (1991)]{Wilson}  Wilson T.L., Johnston K.J.,
        Mauersberger R. 1991,  A\&A 251, 220  
\bibitem[Wood \& Churchwell 1989]{Wood}
	Wood D.O.S., Churchwell E. 1989, ApJSS 69, 831
\bibitem[Young et al. (1998)]{Young} Young L.M., Keto E., Ho P.T.P. 1998, 
        ApJ 507, 270
\bibitem[Zang \& Ho (1997)]{Zang} Zang Q., Ho P.T.P. 1997, ApJ 488, 241



\end{thebibliography}
\end{document}